\documentclass[11pt, oneside]{article}

\usepackage[paper=a4paper, total={16cm,23cm}]{geometry}
\usepackage[T1]{fontenc}
\usepackage{setspace} 
\usepackage[hang]{footmisc} 
\usepackage{titling} 
\usepackage{tocloft} 
\usepackage{parskip} 
\usepackage{comment}

\usepackage{amsfonts}
\usepackage{amsmath}
\usepackage{amssymb}
\usepackage{mathrsfs, dsfont, amsthm}
\usepackage{physics}
\usepackage{mathtools}
\usepackage{upgreek}
\numberwithin{equation}{section} 
\usepackage{wasysym}
\usepackage{marvosym}

\usepackage{xcolor}
\usepackage{colortbl}
\definecolor{dark-red}{rgb}{0.50,0.12,0.12} 
\definecolor{mblue}{rgb}{0.30, 0.45, 0.70}
\definecolor{mred}{rgb}{0.70, 0.20, 0.20}
\definecolor{mgray}{rgb}{0.63, 0.63, 0.63}

\usepackage{hyperref} 
\hypersetup{colorlinks=true, allcolors=dark-red, linktoc=all}

\usepackage{cite}

\usepackage{tikz}
\usepackage{tikzlings}
\usetikzlibrary{calc, patterns, decorations, decorations.markings, shapes, positioning, intersections, quotes, angles}
\usepackage[subrefformat=parens]{subcaption}
\usepackage{pgfplots}
\pgfplotsset{compat=newest}
\newcommand{\mathdefault}[1][]{}

\def \cM {\mathcal{M}}

\newcommand{\beq}{\begin{equation}}
\newcommand{\eeq}{\end{equation}} 

\newcommand{\lc}{\left(}
\newcommand{\rc}{\right)}
\newcommand{\ls}{\left[}
\newcommand{\rs}{\right]} 

\DeclareMathOperator{\arccoth}{arccoth}

\newcommand{\ep}{\mathrm{e}}
\newcommand{\ic}{\mathrm{i}}
\newcommand{\diff}{\mathrm{d}}

\newcommand{\zh}{z_\mathrm{h}}

\begin{document}
\begin{titlingpage}
    \vspace*{3em}
    \onehalfspacing
    \begin{center}
        {\LARGE Imprint of black hole interior on thermal four-point correlators}
    \end{center}
    \singlespacing
    \vspace*{2em}
      \begin{center}
        \textbf{
       Joydeep Chakravarty
        }
    \end{center}
    \vspace*{1em}
    \begin{center}
        \textsl{
        Department of Physics, McGill University \\
        Montr\'eal, QC, Canada \\[\baselineskip]
        }
  \href{mailto:joydeep.chakravarty@mail.mcgill.ca}{\small joydeep.chakravarty@mail.mcgill.ca}
    \end{center}
    \vspace*{3em}
    \begin{abstract}
We consider correlators of light fields smeared against directed wavepackets over a thermal state dual to a single-sided planar AdS black hole.  In the large frequency limit, our measurement is simplified using a bulk WKB description. We propose a dictionary that maps the action of smeared boundary operators to flat-space oscillators near an interior bulk point on the thermal state, by analytically continuing late-time operators from the right to the left boundary via an integral transform. Using the dictionary the smeared correlator factorizes to a flat-space like scattering amplitude about the interior event.  Our transformed correlators describe local physics in the two-sided black hole interior, while incurring a suppression of $\mathcal{O}(e^{-\beta \omega / 2})$.  These measurements necessitate a non-trivial time ordering of operators living on boundary hyperboloids which are causally connected to the past light cone of the bulk point, as well as on a corresponding future branch.    \end{abstract}
\end{titlingpage}
\tableofcontents

\section{Introduction}

The black hole interior is a region of spacetime causally inaccessible from the asymptotic infinity, and is bounded by an event horizon. There are several interesting features associated with the interior region, two prominent examples are an area law for the entropy \cite{Bekenstein-bhae} and black hole evaporation \cite{Hawking-particle-creation}. Recent discussions of the interior have led to remarkable advances in our understanding of quantum gravity, see \cite{Mathur:2009hf, Harlow:2014yka, Almheiri:2020cfm, Raju:2020smc} for recent reviews which capture the progress mostly centred on information-theoretic aspects.

Holography \cite{Maldacena:1997re, Witten-ads-and-holography, Gubser:1998bc} has shed considerable light into our understanding of gravity, and black holes in particular. The statement is quite striking: gravitational physics in an asymptotically anti-de Sitter spacetime is dual to a conformal field theory living on its boundary. This offers a convenient avenue to understand bulk physics while utilizing a non-gravitational description on the boundary. 

Our present objective is to understand what can we learn about thermal correlation functions given that the black hole interior exists. Some early work on thermal correlation functions probing the black hole interior was performed in \cite{Louko:2000tp, Kraus:2002iv}. The thermal one-point function is understood to capture the proper time to the singularity \cite{Grinberg:2020fdj}, see \cite{David:2022nfn, Horowitz:2023ury ,David:2023uya, Singhi:2024sdr} exploring different sub-cases. It is also well-established that analytic continuation of the two-point function to the second sheet gets contributions from bouncing geodesics reflected off the black hole singularity \cite{Fidkowski:2003nf, Festuccia:2005pi, Festuccia:2008zx} (see also \cite{Amado:2008hw}). Using this language, the singularity was also recently studied using the OPE \cite{Ceplak:2024bja, Ceplak:2025dds} along with stringy effects \cite{Dodelson:2025jff}. Recently it has been understood that the spectral density at a finite-temperature in a holographic theory is a simple natural observable which receives contributions from summing over geodesics reflected off both the past and the future black hole singularities \cite{Afkhami-Jeddi:2025wra}.  

\begin{figure}[t!]
    \centering
        \begin{subfigure}{0.8\textwidth}
    \centering
       \begin{tikzpicture}[scale=0.8]
    \draw[color=mblue, decoration={markings,mark=at position 0.5 with \arrow{>}},postaction=decorate,thick] (-2,-0.5) -- (-3.5, 1) arc (225:135:.05) coordinate (turning);
   \draw[dashed] (turning)+(-1,-1)-- ++(0.95,0.95); 
     \draw[color=mblue!50, decoration={markings,mark=at position 0.75 with \arrow{>}},postaction=decorate, thick] (turning) arc (135:45:.05) -- ++(1.40,-1.40) ++(0.2,-0.2)--(-.24,-2.13);
    \draw[color=mblue,decoration={markings,mark=at position 0.7 with \arrow{<}},postaction=decorate, thick] (0,2.12) -- (-2,-0.5);
     \draw[color=mblue,decoration={markings,mark=at position 0.25 with \arrow{>}},postaction=decorate, thick] (-2.2,-2.62) -- (-2,-0.5);
     \draw[color=mblue,decoration={markings,mark=at position 0.5 with \arrow{>}},postaction=decorate, thick] (-1.4,-2.32) -- (-2,-0.5);
     \draw (-1.7,-0.5) node[anchor = west]{$X$};
     \draw (0, 2.12) node[anchor = south]{$O_3$};
     \draw (-.24,-2.13) node[anchor = north]{$O_4$}; 
    \draw (-2.2,-2.62) node[anchor = north]{$O_1^\dagger$}; 
    \draw (-1.4,-2.32) node[anchor = north]{$O_2^\dagger$}; 
    \draw[thick] (-3,3) to[out=-30,in=-150] (3,3) node[anchor = north west]{$H^+(X)$};
    \draw[thick] (-3,-3) to[out=30,in=150] (3,-3) node[anchor = south west]{$H^-(X)$};
        \draw[->] (5,0)--(6,0);
    \draw[->] (5,0)--(5,1);
    \draw[->] (5,0)--(4.4,-0.4);
    \node [right] at (6,0) {$\bf{x}$}; 
    \node [above] at (5,1) {$t$}; 
    \node [below] at (4.1,-0.4) {$z$}; 
    \end{tikzpicture}
    \caption{}
\end{subfigure}

    \begin{subfigure}{0.45\textwidth}
    \centering
     \begin{tikzpicture}[scale=0.8] 
\draw  (0,0.1)--(0,0) -- (5.5,0) -- (5.5,-0.3) -- (0,-0.3)--(0,-0.4);
\node [below] at (1.2,-0.3) {$O_{x_4, p_4, \sigma_4}$};
\node [above] at (4.5,-0) {$O_{y_3, q_3, \sigma_3}$};
\node [above] at (0.6,0) {$O^\dagger_{x_1, p_1, \sigma_1}$};
\node [above] at (2.4,-0) {$O^\dagger_{x_2, p_2, \sigma_2}$};
\filldraw (4.5,0) circle (2pt); \filldraw (1.5,-0.3) circle (2pt); \filldraw (1,0) circle (2pt); \filldraw (2,0) circle (2pt);
\end{tikzpicture}
\caption{}
\end{subfigure}
    \begin{subfigure}{0.45\textwidth}
    \centering
     \begin{tikzpicture}[scale=0.8] 
\draw  (0,0.1)--(0,0) -- (5.5,0) -- (5.5,-0.3) -- (0,-0.3)--(0,-0.4);
\node [below] at (1.2,-0.3) {$a_{X, P_4}$};
\node [above] at (4.5,-0) {$b_{X, Q_3}$};
\node [above] at (0.6,0) {$a^\dagger_{X, P_1}$};
\node [above] at (2.4,-0) {$a^\dagger_{X, P_2}$};
\filldraw (4.5,0) circle (2pt); \filldraw (1.5,-0.3) circle (2pt); \filldraw (1,0) circle (2pt); \filldraw (2,0) circle (2pt);
\end{tikzpicture}
\caption{}
\end{subfigure}
    \caption{(a) Radar scattering process  $\langle \Psi |  \, O_{x_4, p_4, \sigma_4} \, O_{y_3, q_3, \sigma_3}\,  O^\dagger_{x_2, p_2, \sigma_2} \, O^\dagger_{x_1, p_1, \sigma_1} \,  | \Psi \rangle$ with a bulk point in black hole exterior as in \cite{Caron-Huot:2025hmk, Caron-Huot:2025she}, see equation \eqref{eq:bdryops} for the definition of smeared operators. The dashed line is trace over all final states that may fall behind the horizon. The process has three early-time (two emission and one absorption) operators  and a single late-time (absorption) operator. (b) The radar correlator on an in-in timefold. (c) The bulk scattering amplitude with same time-ordering. The left ends of (b) and (c) incorporate the thermal density matrix, which technically is a circle in Euclidean time, but to keep notation light, we do not indicate it and use the same notation as a state. }
    \label{fig:extbpcontour}
\end{figure}
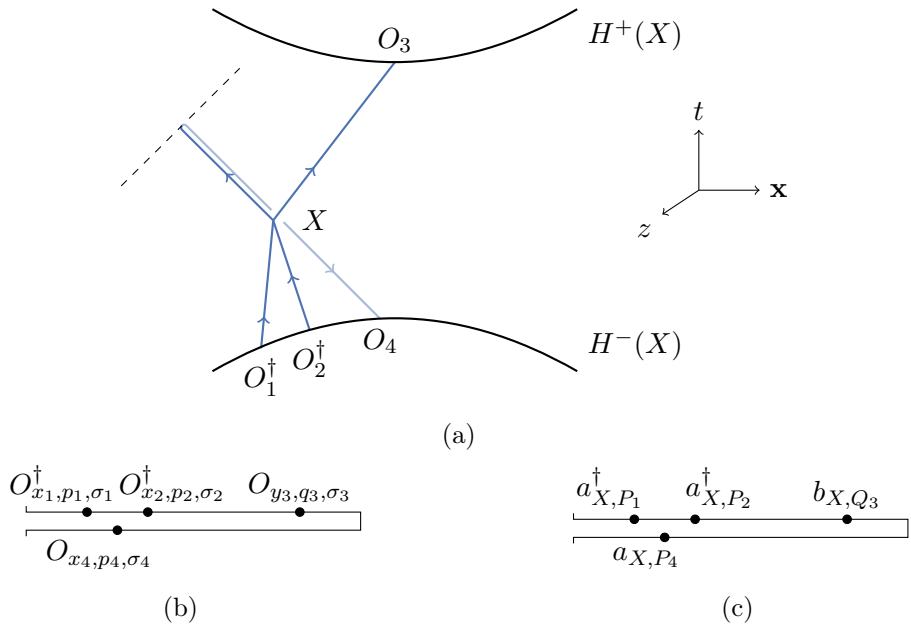
Our present interest is to understand whether we can learn more using higher point correlators. At four and higher points, boundary correlation functions can be used to address bulk locality. It is understood that boundary correlators can be tuned to capture bulk scattering processes by extracting flat-space physics \cite{Polchinski:1999ry, Giddings:1999jq, Gary:2009ae, Okuda:2010ym, Penedones:2010ue}. This provides us with a crisp signature of bulk locality as clearly outlined in \cite{Maldacena:2015iua}.

We will utilize well-posed exterior measurements on $k$-point correlators involving bulk locality to image the black hole interior, and thus study its imprint on boundary thermal correlation functions\footnote{Here we share the same theme as for two-point functions in \cite{Afkhami-Jeddi:2025wra} where the Fourier transform of the spectral density (i.e. a measurement involving the single-sided black hole exterior) picks up an analytically continued path and thus receives contributions from reflected geodesics in the two-sided geometry. See Fig 3 of \cite{Afkhami-Jeddi:2025wra}. } \footnote{For a generic correlator, our interior description holds for $k$-point correlators with $k \ll \mathcal N$ upto $\mathcal O (e^{- \mathcal N})$ corrections. We do not expect the interior description to hold beyond this as physically we can argue that bulk locality breaks down if we perform a complicated measurement involving a $\mathcal O ( \mathcal N)$ operator insertions.}. The flat space limit can also be taken in general geometries \cite{Caron-Huot:2025hmk, Caron-Huot:2025she}. These works propose a dictionary between local creation-annihilation operators about the bulk point to boundary operators boundary correlators. Using this thermal correlators factorize to flat-space like amplitudes in the almost-flat neighborhood about a bulk point. The boundary signature of these processes is imprinted on \textit{boundary hyperboloids}, which are surfaces causally connected to the bulk point \cite{Caron-Huot:2025she}.

\begin{figure}[t!]
 \begin{subfigure}{1.0\textwidth}
    \centering
     \begin{tikzpicture}[scale=0.6]
 \node [right] at (0,-0.15) {$\quad$}; 
\draw  (0,0.1)--(0,0) -- (5,0) -- (5,-0.3) -- (0,-0.3)--(0,-0.4);
\node [above] at (4,0) {$O(y_1)$};
\filldraw (4,0) circle (2pt);
\node [right] at (6.5,-0.2) { $  \mapsto $};
\draw  (9,0.1)--(9,0) ..controls (11,0.1) and (12,1.1).. (13.5,1.0) -- (13.5,-0.3) -- (9,-0.3)--(9,-0.4);
\node [above] at (12.5,1.0) {$O_L(y_1)$};
\node [right] at (15.3,0.25) {\small ${\beta \over 2}$};
\filldraw (12.7,0.95) circle (2pt); 
\draw[<->] (15, 1)-- (15, -0.3);
\draw[mred] [-{latex[scale=5]}](12.7,0.0) -- (12.7,0.75);
\end{tikzpicture}
\caption{}
\end{subfigure}
 \begin{subfigure}{1.0\textwidth}
    \centering
     \begin{tikzpicture}[scale=0.6]
 \node [right] at (0,-0.15) {$\quad$}; 
\draw  (0,0.1)--(0,0) -- (5,0) -- (5,-0.3) -- (0,-0.3)--(0,-0.4);
\node [below] at (4,-0.3) {$O(y_2)$};
\filldraw (4,-0.3) circle (2pt);
\node [right] at (6.5,-0.2) { $ \mapsto $};
\draw  (9,0.1)--(9,0) -- (13.5,0) -- (13.5,-1.3) ..controls (12,-1.4) and (11,-0.2).. (9,-0.3)--(9,-0.4);
\node [below] at (12.5,-1.3) {$O_L(y_2)$};
 \node [right] at (15.3,-0.55) {\small ${\beta \over 2}$};
 \filldraw (12.7,-1.25) circle (2pt);
\draw[<->] (15, 0)-- (15, -1.3);
\draw[mred] [-{latex[scale=5]}](12.7,-0.3) -- (12.7,-1.05);
\end{tikzpicture}
\caption{}
\end{subfigure}
\begin{subfigure}{1.0\textwidth}
    \centering
\begin{tikzpicture}[scale=0.85] 
\draw  (0,0.1)--(0,0) -- (5.5,0) -- (5.5,-0.3) -- (0,-0.3)--(0,-0.4);
\node [below] at (1.2,-0.3) {$O_{x_4, p_4, \sigma_4}$};
\node [above] at (4.5,-0) {$O_{y_3, q_3, \sigma_3}$};
\node [above] at (0.6,0) {$O^\dagger_{x_1, p_1, \sigma_1}$};
\node [above] at (2.4,-0) {$O^\dagger_{x_2, p_2, \sigma_2}$};
\filldraw (4.5,0) circle (2pt); \filldraw (1.5,-0.3) circle (2pt); \filldraw (1,0) circle (2pt); \filldraw (2,0) circle (2pt);
\node [right] at (6.5,-0.2) { $  \mapsto $};
\draw  (8,0.1)--(8,0) ..controls (12,0.1) and (12.5,1.1).. (13.5,1.0) -- (13.5,-0.3) -- (8,-0.3)--(8,-0.4);
\node [below] at (9,-0.3) {$O_{x_4, p_4, \sigma_4}$};
\node [above] at (12.5,0.95) {$O^\dagger_{Ly_3, q_3, \sigma_3}$};
\node [above] at (8.2,0.2) {$O^\dagger_{x_1, p_1, \sigma_1}$};
\node [above] at (10.0,0.3) {$O^\dagger_{x_2, p_2, \sigma_2}$};
\filldraw (13,0.95) circle (2pt); \filldraw (8.7,-0.3) circle (2pt); \filldraw (8.3,0.04) circle (2pt); \filldraw (9.2,0.05) circle (2pt);
\end{tikzpicture}
\caption{}
\end{subfigure}
    \caption{(a) and (b) On a Schwinger Keldysh (SK) fold with a thermal identification, the transform \eqref{transformaton} takes the high frequency operator insertions $y$ on the right future boundary as operators introduced at the left boundary. There is an energy cost for both operations of $\mathcal{O}(e^{-\beta \omega / 2})$, which we absorb in our operational definition \eqref{finform}. While the transform is a robust operation, geometrically these operations on saddles violate path integral rules (for instance KSW criteria in the bulk \cite{Kontsevich:2021dmb, Witten:2021nzp}). (c) The transformation in \eqref{finform} takes us from an exterior radar experiment defined on the SK fold to a deformed SK fold, thereby implementing $t_3  \mapsto -t_3 + {\ic \beta / 2}$ such that $O_{y_3, q_3, \sigma_3}$ is shifted to $O^\dagger_{Ly_3, q_3, \sigma_3}$. }
    \label{fig:ancotr}
\end{figure}
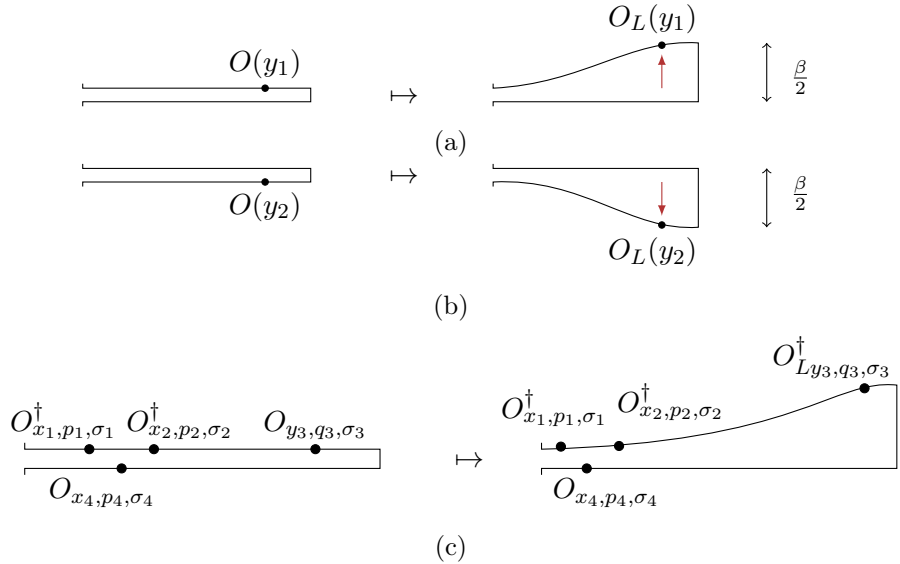

We now employ similar tools to understand the black hole interior using boundary thermal correlation functions. In our work we specifically explore the following questions: 
\begin{enumerate}
\item Can suitable exterior experiments described using thermal correlators be tuned to capture local physics in an analytically continued interior region?
\item What is the signature of scattering processes localized in a flat-space neighborhood about a bulk point within the black hole interior?
\item More precisely, do we have a (state-dependent) dictionary which maps local creation-annihilation operators about the interior bulk point to boundary operators?
\item In light of the above questions, what does the existence of the black hole interior tell us about thermal correlation functions?
\end{enumerate}

We build upon the ground work developed on the lines of \cite{Caron-Huot:2025hmk, Caron-Huot:2025she}. We work with directed wavepackets (see also \cite{Polchinski:1999ry, Chandorkar:2021viw, Terashima:2023mcr}) and inclusive amplitudes (see \cite{Kosower:2018adc, Caron-Huot:2023vxl, Caron-Huot:2023ikn, Aoude:2024sve} for generalised amplitudes). Holographic cameras, which are a coincidence limit of these correlators, also employed both these aspects to study bulk locality \cite{Caron-Huot:2022lff}. Smeared Fourier transforms on the boundary operators within the conformal correlator give rise to directed wavepackets in the bulk specified as a function of the boundary shooting momentum $p$. This allows us to locate the bulk point $X$ in general geometries.

The second ingredient is non-trivial time-ordering within conformal correlators (see \cite{Hartman:2015lfa, Maldacena:2015waa}). This allows us to capture a variety of physical processes as we can also have early time absorption operators and late time emission operators inserted within the correlator. Our present interest is to obtain these bulk scattering amplitudes using boundary correlators. We also discuss generalizations to scattering amplitudes on two-timefolds.

 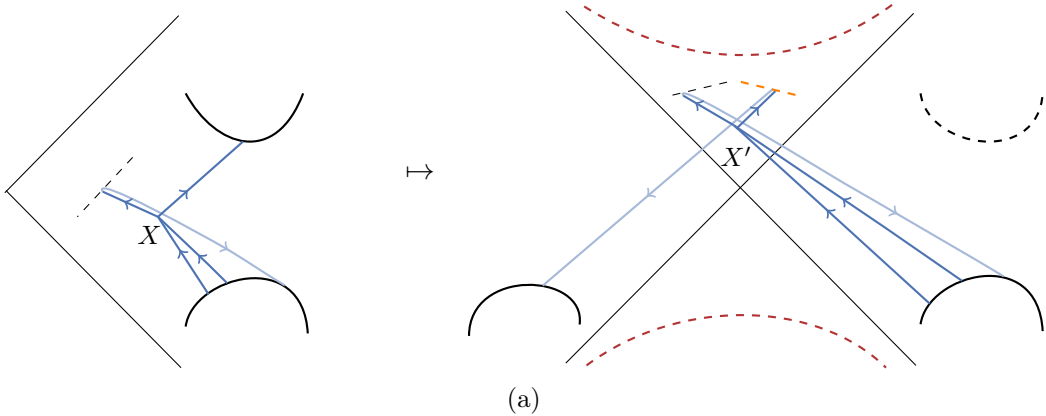
\begin{figure}[t!]
 \begin{subfigure}{1.0\textwidth}
    \centering
\begin{tikzpicture}[x=0.75pt,y=0.75pt,yscale=-1.15,xscale=1.15]
\draw    (176,185.5) -- (252,263) ;
\draw    (251,109) -- (175.5,186.5) ;
\draw [thick]   (254,245) .. controls (257,223) and (305,210) .. (307,248) ;
\draw  [thick]  (254,143) .. controls (267,167) and (293,175) .. (305,143) ;
\draw [thick, color=mblue  ,decoration={markings,mark=at position 0.5 with \arrow{<}},postaction=decorate ,draw opacity=5 ]   (242,197) -- (264,231) ;
\draw [thick, color=mblue  ,decoration={markings,mark=at position 0.65 with \arrow{<}}, postaction=decorate  ,draw opacity=5 ]   (242,197) -- (264,218) -- (272,226) ;
\draw [thick, color=mblue!50  ,decoration={markings,mark=at position 0.35 with \arrow{<}}, postaction=decorate  ,  ,draw opacity=5 ]   (297,227) .. controls (287.16,220.95) and (277,214) .. (270,210) .. controls (263,206) and (213.13,177.89) .. (218,186) ;
\draw [thick, color=mblue  ,decoration={markings,mark=at position 0.35 with \arrow{>}}, postaction=decorate  ,draw opacity=5 ]   (242,197) -- (261.67,179.45) -- (279,164) ;
\draw [thick, color=mblue  ,decoration={markings,mark=at position 0.5 with \arrow{<}}, postaction=decorate  ,draw opacity=5 ]   (218,186) -- (242,197) ;
\draw[dashed]    (231,171) -- (207,197) ;
\node [right] at (345,178.1)(6.5,-0.2) { $  \mapsto $};
\draw (232,200) node [anchor=north west][inner sep=0.75pt]  [font=\small] [align=left] {$X$};
\draw    (419,107) -- (571,262) ;
\draw    (570,108) -- (419,261) ;
\draw  [thick]  (573,244) .. controls (576,222) and (624,209) .. (626,247) ;
\draw[thick]    (425,244.1) .. controls (428,221.2) and (377,219.1) .. (377,249.1) ;
\draw  [dashed, thick]  (573,143) .. controls (576,167) and (624,175) .. (626,143) ;
\draw [thick, color=mblue  ,decoration={markings,mark=at position 0.5 with \arrow{<}}, postaction=decorate  ,draw opacity=5 ]   (492,157) -- (577,235) ;
\draw [thick, color=mblue  ,decoration={markings,mark=at position 0.5 with \arrow{<}}, postaction=decorate  ,draw opacity=5 ]   (492,157) -- (558,202) -- (591,225) ;
\draw [thick, color=mblue!50  ,decoration={markings,mark=at position 0.35 with \arrow{<}}, postaction=decorate  ,draw opacity=5 ]   (609,223) .. controls (599.16,216.95) and (548,187) .. (541,183) .. controls (534,179) and (465.13,135.89) .. (470,144) ;
\draw [thick, color=mblue!50  ,decoration={markings,mark=at position 0.55 with \arrow{>}}, postaction=decorate  ,draw opacity=5 ] (509,141)-- (409,227.1) ;
\draw [thick, color=mblue  ,decoration={markings,mark=at position 0.55 with \arrow{<}}, postaction=decorate  ,draw opacity=5 ] (510,142)--(493,158.5);
\draw [thick, dashed, mred ,draw opacity=5 ]   (427,105) .. controls (471,133) and (521,134) .. (561,104) ;
\draw [thick, dashed, mred  ,draw opacity=5 ]   (428,262) .. controls (468,232) and (528,233) .. (558,263) ;
\draw [thick, color=mblue,decoration={markings,mark=at position 0.35 with \arrow{<}}, postaction=decorate   ,draw opacity=5 ]   (470,144) -- (492,157) ;
\draw[dashed]    (489,138.1) -- (464,144.1);
\draw[dashed, thick, orange ](495,138) -- (520,144);
\draw (485,165.1) node [anchor=north west][inner sep=0.75pt]  [font=\small] [align=left] {$X'$};
\end{tikzpicture}
\caption{}
\end{subfigure}

 \caption{The analytic continuation can be thought of as pushing the bulk point $X$ from a well-defined exterior scattering experiment on single timefold in Fig \ref{fig:extbpcontour} to an interior scattering experiment with bulk point $X'$. The transformation introduces an extra folding (dashed orange line), and local right-moving modes near $X'$. Note that the boundary hyperboloids for exterior bulk point experiment and the interior bulk point experiment are different, and we need to change the operator positions and the shooting momenta appropriately to go from $X$ to $X'$.}
    \label{mainfig00}
\end{figure}


\subsection*{Methods and results}
We address the following question:  
can one push the bulk point to the black hole interior?
We start with a well-defined scattering process about a bulk point in the exterior, and perform an analytic continuation to another process whose bulk point lies in the interior. Consider the following transform acting on an right operator $O_R(t')$ which defines an operator on the left:
\beq \label{transformaton}
{\beta \over 2\pi} \int_{-\infty}^\infty  {{\rm d} t' \over (t'+t)^2 +{\beta^2\over 4}} \ e^{\pm \ic \omega t'} \ O_R(t') = e^{-\beta \omega \over 2} \ e^{\mp \ic \omega t} \ O_R\lc -t \pm \frac{\ic \beta}{2} \rc
\eeq
We direct the reader to the discussion to \S \ref{sec:lsnbp} and especially to equations \eqref{finform} and \eqref{eq: smearing function} for the precise implementation. As shown in Fig \ref{fig:ancotr}, this localizes the operator insertions to the left boundary since $O_R( -t \pm \frac{\ic \beta}{2} ) \equiv O_L(t)$, while continuing the positive and the negative frequency modes differently. 
Using wavepackets, we use these operators to obtain local right-moving modes in the interior. 
Since in the high energy limit the bulk-to-boundary propagator localizes to a geodesic description, the resulting description can be extended straightforwardly into the interior.

\paragraph{Experiments:}
We are interested in two interior experiments: the radar experiment in Fig \ref{mainfig00} whose bulk picture is in Fig \ref{fig:ancotr}, and the otoc experiment using Fig \ref{fig:otocexp} whose bulk picture is in Fig \ref{fig:2to2int}. In both cases we analytically continue a $4$-pt correlator defined on timefolds from the exterior to the black hole interior. 

We work in the probe limit where we set $N=\infty$. Note that analytic continuation to the interior requires that the exterior experiment has a large enough time separation between initial and final operator insertions $\Delta t = \abs{t_{i} - t_{f}}$. Hence we obtain the bulk point beyond a corresponding depth in the interior. 

\paragraph{Radar experiment:} Let us briefly summarize the experiment, where the interior radar correlator can be written as:
\beq
\begin{split}
\langle \Psi |  \, O_{x_4, p_4, \sigma_4} \, O_{Ly_3, q_3, \sigma_3}\, & O^\dagger_{x_2, p_2, \sigma_2} \, O^\dagger_{x_1, p_1, \sigma_1} \,  | \Psi \rangle \\
& = \sum_{\rm out} \langle \Psi   |\, O_{x_4, p_4, \sigma_4} \,|  {\rm out} \rangle \, \langle {\rm out}  |\, O_{Ly_3, q_3, \sigma_3} \,  O^\dagger_{x_2, p_2, \sigma_2} \, O^\dagger_{x_1, p_1, \sigma_1} \, | \Psi \rangle
\end{split}
\eeq
The sum over the out states is represented by the dashed line in Fig \ref{mainfig00}(a). Consequently we do not need to keep track of the final microstate after the experiment. The bulk-to-boundary propagators at high-energies reduce to simple phases near the bulk point:
\begin{equation*}
\begin{split}
&\expval{\, \Phi(X +\delta X) \, O_{x,p,\sigma}^\dagger \, }{\Psi} \propto \expval{\, \Phi(X +\delta X) \, a^\dagger_{X,P} \, }{\Omega} =\ep^{\ic P \cdot \updelta X}, \qquad ({\rm plane \ wave \ limit, \ high \ energy}) 
\end{split}
\end{equation*}
and similarly for the annihilation modes with  phase $\ep^{-\ic P \cdot \updelta X}$.
Using the smeared operators $O_{Ly,q,\sigma}^\dagger / O_{Ly,q,\sigma}$ on the left boundary we can similarly extract the \textit{late-time} phases $\ep^{\pm \ic Q \cdot \updelta X}$. 
Extracting the {early-time} and the {late-time} phases associated with the plane-wave expansion about the bulk-point enables us to describe flat-space processes in the black hole interior. We provide this operator dictionary specified by its action on the state in Table \ref{movers} (in the plane wave limit where expressions simplify).

Consequently the radar correlator factorizes in terms of a $3 \to 1$ flat space like amplitude about an interior bulk point $X$. Using the phases, we can obtain a factorization formula for the correlator in terms of a local flat space like scattering amplitude \cite{Caron-Huot:2025hmk}. 

\begin{table}[t!]
    \centering
    \renewcommand{\arraystretch}{1.5}
    \setlength{\tabcolsep}{12pt}
    \arrayrulecolor{gray}
    \begin{tabular}{|c|c|c|c|c|}
    \hline
    Region &Type& Absorption modes & Emission modes \\ 
    & &rhs  $\times \lc \sqrt{\mathcal{D}(p; X)} \, e^{\ic S(p; X)}\rc $ & rhs $\times \lc \sqrt{\mathcal{D}(p; X)} \, e^{\ic S(p; X)}  \rc  $\\ \hline
    Exterior & Left movers & ${O}_{x,p, \sigma} | \Psi  \rangle  \; \to \; \mathcal{C}^-_{\Delta,p} a_{X,{P}}|\Omega \rangle$  & ${O}^\dagger_{x,p,\sigma} | \Psi  \rangle\; \to \;   \mathcal{C}^+_{\Delta,p} a^\dagger_{X,{ P}}|\Omega \rangle$ \\ 
    & Right movers & ${O}_{y,q, \sigma} | \Psi  \rangle\; \to \;   \mathcal{C}^-_{\Delta,q} b_{X,{ Q}}|\Omega \rangle$ & ${O}^\dagger_{y,q,\sigma}| \Psi  \rangle \; \to \;   \mathcal{C}^+_{\Delta,q} b^\dagger_{X,{ Q}}|\Omega \rangle$  \\ \hline
    Interior & Left movers & ${O}_{x,p,\sigma} | \Psi  \rangle \; \to \;   \mathcal{C}^-_{\Delta,p} a_{X,{ P}}|\Omega \rangle$  & ${O}^\dagger_{ x,p, \sigma} | \Psi  \rangle\; \to \;   \mathcal{C}^+_{\Delta,p} a^\dagger_{X,{ P}}|\Omega \rangle$ \\ 
     & Right movers & ${O}_{Ly,q, \sigma} | \Psi  \rangle\; \to \;   \mathcal{C}^-_{\Delta,q} \tilde{a}^\dagger_{Y,{ Q}}|\Omega \rangle$ & ${O}^\dagger_{Ly,q, \sigma}| \Psi  \rangle \; \to \;   \mathcal{C}^+_{\Delta,q} \tilde{a}_{Y,{ Q}}|\Omega \rangle$  \\ \hline
\end{tabular}
\caption{The plane-wave limit $\lim_{\sigma \to \infty} {O}_{x,p,\sigma}$ of the dictionary defined by their action on the state, between the bulk and boundary modes in the black hole geometry. Here $x$ and $y$ denote the past and the future boundary hyperboloids, while $x$ and $Ly$ denotes the hyperboloids on the right and the left CFTs, which are connected to the bulk point by the left and the right movers respectively. We overcome the exponential suppression $\mathcal{O}(e^{-\beta \omega / 2})$ by boosting the transformation appropriately using \eqref{eq: smearing function}.}\label{movers}
\end{table}

\paragraph{Boundary locus:} The lightcones emanating from the bulk point lead to spacelike curves on the boundary known as boundary hyperboloids (related to lightcone cuts of \cite{Engelhardt:2016wgb, Engelhardt:2016crc}). As shown in \cite{Caron-Huot:2025she}, the physics of these curves are captured by the Hamilton-Jacobi function $S(p,X)$. The variational principle also allows us to capture features of bulk variations in terms of boundary variations along the boundary hyperboloids. 

We show that the boundary hyperboloids have interesting shapes corresponding to different locations of the bulk point in the black hole interior. Operations upon these hyperboloids allow us to understand different aspects of an infalling observer. We find here that the hyperboloids reveal a distinct visual signature as we approach the black hole singularity.

\paragraph{Connections to literature:} Looking at thermal correlators in holography naturally leads us to connections with other important areas, we list a few important references where they have played a huge role. Following the original proposal of \cite{Iliesiu:2018fao}, there has been some exciting recent progress on analytic thermal bootstrap technology for holographic correlators by trying to implement KMS properly, see the works \cite{Esper:2023jeq, Buric:2025fye, Buric:2025anb} and \cite{Barrat:2025nvu, Barrat:2025twb, Barrat:2025wbi}. This very nicely complements recent interesting developments in understanding different aspects of finite temperature thermal correlators computed holographically using bulk description \cite{Alday:2020eua, Loganayagam:2025ell, Krishna:2021fus, Loganayagam:2022teq} (see also \cite{Rodriguez-Gomez:2021pfh, Georgiou:2022ekc, Georgiou:2023xpg}). Related developments also include the thermal product formula \cite{Dodelson:2023vrw, Bhattacharya:2025vyi}, bootstrapping by modelling OPE tails \cite{Niarchos:2025cdg}, an improved understanding of analyticity \cite{Loganayagam:2022teq, Banerjee:2024dpl, Das:2024fwg} and another interesting thermal observable \cite{David:2025tqn}.

Finally, the factorization formula \eqref{rep2} and its extension to the interior generalize various flat space limits of AdS/CFT obtained about the vacuum AdS case. Some important avenues where flat-space limits in holography (and with similar motivations) have been discussed are in the context of gapped operator spectrum \cite{Heemskerk:2009pn, Fitzpatrick:2010zm}, Mellin amplitudes \cite{Fitzpatrick:2011ia, Goncalves:2014ffa}, total energy pole \cite{Raju:2012zr}, Landau diagrams in AdS \cite{Komatsu:2020sag}, by putting sharp CFT bounds \cite{Caron-Huot:2021enk}, using HKLL reconstruction \cite{Hijano:2019qmi, Duary:2022pyv}, the Regge-growth conjecture \cite{Chandorkar:2021viw}, boundary formulation of on-shell  S-matrix \cite{Jain:2023fxc}, AdS radius corrections to S-matrix \cite{Banerjee:2022oll}, massive particles \cite{Gadde:2022ghy}, flat-space partial waves from OPE \cite{vanRees:2023fcf}, celestial and Carrollian bulk-point kinematics \cite{deGioia:2024yne, Alday:2024yyj}. We direct the reader to \cite{Li:2021snj} for a review of the subject and for further references.

\section{Flat-space physics about exterior bulk point}
\label{sumexp}

We will recap the important features introduced in \cite{Caron-Huot:2025hmk, Caron-Huot:2025she} which are necessary to locate a bulk point. Bulk points in the exterior region can be located using either in-out correlators or inclusive correlators on timefold contours.  
We discuss the significance of these observables in \S \ref{incobv4pt}, which would be useful to generalize to the interior discussion.

\subsection{Zooming near a bulk point using WKB approximation}
Let us define the smeared operators lying on the boundary hyperboloids:
\begin{equation}\label{eq:bdryops}
\begin{split}
    O_{x^A,p^A,\sigma^A} &= \int {\rm d}^d \qty(\delta x) \, \psi^*_{p^A,\sigma^A}(\delta x) \, O(x^A + \delta x), \\ 
    O^\dagger_{x^A,p^A,\sigma^A} &= \int {\rm d}^{d} \qty(\delta x) \, \psi_{p^A,\sigma^A}(\delta x) \, O(x^A + \delta x),
\end{split}
\end{equation}
where the label $A$ runs over the past and the future hyperboloids, i.e. shooting wavepackets from past hyperboloid $x$ with shooting momentum $p$, and from the future hyperboloid $y$ with shooting momentum $q$ respectively. Here we have defined the smeared Fourier transform as given below:
\begin{equation} \label{psips}
    \psi_{p,\sigma}(\delta x) =
    \exp(\ic p_\mu \delta x^\mu - \frac{1}{2} \sigma^{-1}_{\mu\nu} \, \delta x^\mu \delta x^\nu), \qquad \det \sigma > 0,
\end{equation}
The smeared operators act upon the boundary state $\ket{\Psi}$ which is dual to the planar black hole geometry described by the following metric:
\begin{equation}\label{eq:pbhmetric}
    \diff s^2 = \frac{1}{z^2} \qty(-f(z) \, \diff T^2 + \frac{\diff z^2}{f(z)}+ \delta_{ab} \, \diff X^a \diff X^b ), \qquad f(z) =1 - \frac{z^d}{\zh^d}.
\end{equation}
The inverse Hawking temperature of this black hole is given by $\beta = {4 \pi z_{\rm h} / d}$.
In these coordinates, the black hole horizon is at $z = \zh$, while $z< \zh$ covers the black hole exterior and the conformal boundary lies at $z =0$. 

 \subsubsection{Propagator and geodesics}
 The boundary correlator can be decomposed into product of bulk-to-boundary propagators which are our observables of interest. As in \cite{Caron-Huot:2025hmk}, we work with the symmetrized Wightman function. 
\beq
    \Phi_{x,p,\sigma}(X) = \expval{\, \Phi(X) \, O_{x, p,\sigma}^\dagger \, }{\Psi}
\eeq
where $| \Psi \rangle$ is a CFT state dual to the black hole geometry in \eqref{eq:pbhmetric}. The processes we consider correspond to large energies, i.e. $\beta \omega_i \gg 1$. 

The bulk equation of motion in general geometries has a Heun form and is given in \eqref{eq:eom-scalar-pbh}. To extract a clean answer from this, we work with a WKB approximation of this solution as specified in \cite{Caron-Huot:2025hmk}. This takes the answer of the form in \eqref{sol1}. Since the propagator also solves the bulk equation of motion, using the bulk solution \eqref{sol1} the propagator takes the form of a simple phase in the large radial momentum limit:
\begin{equation}\label{eq:Phi X with normalization}
    \expval{\Phi(X) \, O_{p}^\dagger }{\Psi} = \mathcal{C}^\pm_{\Delta,p} \sqrt{\mathcal{D}(p; X)} \exp(\ic S(p; X)),
\end{equation}
The normalization constant is given in \eqref{normalization1}. The notation in \eqref{eq:Phi X with normalization} is explained as follows.
\begin{enumerate}
\item The exponential phase $S(p,X)$ is the first-order WKB contribution which is a canonical transformation of the Hamilton-Jacobi function $S(x;X)$ \cite{Caron-Huot:2025she}.
\beq \label{HJ-function}
S(p,X) = p\cdot x(p,X) + S\lc x(p,X);X\rc \quad 
\eeq
This is essentially the geometric optics regime. Using \eqref{sol1}, we obtain the on-shell action corresponding to the geodesic.
\beq
S(x;X) = {1 \over 2 } \int_x^X \diff \sigma \lc {g_{MN} \over \eta} {\diff X^M \over \diff \sigma} {\diff X^N \over \diff \sigma} -\eta m^2 \rc
\eeq
A small variation about the bulk point $X$ is recorded using the Hamilton-Jacobi function and obeys the following simple relation:
\beq
   \exp(\ic S(p; X + \updelta X)) = \exp(\ic S(p; X) + \ic P_M \updelta X^M + \mathcal{O}(\updelta X^2)).
\eeq
At the level of geometric optics the Fourier transform defined on the boundary operators in \eqref{eq:bdryops} implements the canonical transformation in \eqref{HJ-function}. 

\item The pre-factor $\mathcal{D}(p; X) $ arises from the second-order WKB approximation and characterizes physical optics contributions. From the perspective of the path integral, it is understood as the van Vleck-Morette determinant \cite{VanVleck:1928zz, Morette:1951zz, Visser:1992pz}. This is essentially the one-loop determinant which evaluates to the following value for the black hole geometry.
\beq \label{vvmdet}
\mathcal{D}(p; X)= \frac{z^{d-1}}{2f(z)P_z}
\eeq
Here we have labelled the radial momentum  as $P_z$ which is given below:
\beq
P_z = \sqrt{- \frac{g^{\mu\nu}}{g^{zz}} p_\mu p_\nu} = {1 \over f(z)} \sqrt{{E^2 - \mathbf{p}^2 f(z)}}
\eeq 
such that the bulk momentum corresponding to the geometry \eqref{eq:pbhmetric} is given by $P = \lc p_\mu, P_z \rc$.
\end{enumerate}

 Orbits in black hole backgrounds using a similar WKB have been used in \cite{Dodelson:2022eiz} (see also \cite{Berenstein:2020vlp} and \cite{Chen:2025cee, Chen:2025ywj, Baishya:2024sym} for some detailed analysis in different cases), whereas bulk light-cone singularities have been described in \cite{Dodelson:2023nnr}.

\subsubsection{Dictionary between local flat space modes and boundary operators}
In the large radial momentum limit, since the WKB approximation reduces the problem to a geodesic description, we can single out a bulk point $X$ by tuning the boundary shooting momentum $p$. 

Using this local mode expansion near the bulk point, the quantum field can be expressed using a basis of early-time modes $a_{X,P} /a^\dagger_{X,P}$ or a basis of late-time modes $b_{X,P} /b^\dagger_{X,P}$. This allows us to relate the boundary smeared operators as local bulk operators probing the local flat-space vacuum $\ket{\Omega}$. 
\begin{equation}\label{eq:dictionarysmeared}
  O^\dagger_{p,\sigma} \ket{\Psi } \;\simeq\; 
    \begin{cases}
        \int \ls {\rm d} P \, \psi^+ \rs \ a^\dagger_{X,P}  \ket{\Omega}, & \mbox{if $O^\dagger$ is in the past of $X$}, \\
        \int \ls {\rm d} P \, \psi^- \rs \ b^\dagger_{X,P} \ket{\Omega}, & \mbox{if $O^\dagger$ is in the future of $X$}.
    \end{cases}
\end{equation}
The two cases arise since we consider the Wightman function. Here the notation for the measure is given in \eqref{prodpsi} and the bulk wavepacket $\psi^\pm(P)$ is given in \eqref{psi bulk good}.  The above causal relations imply that indicating the hyperboloid completely fixes the oscillator mode.

The plane wave limit of this phase expansion \eqref{eq:Phi X with normalization} allows us to relate the bulk modes with the boundary smeared operators in a particularly simple form.
\begin{equation}\label{eq:dictionary}
  O^\dagger_{p} \;\simeq\;  \sqrt{\mathcal{D}(p; X)} \, e^{\ic S(p; X)} 
    \times
    \begin{cases}
        \mathcal{C}^+_{\Delta,p}a^\dagger_{X,P}, & \mbox{if $O^\dagger$ is in the past of $X$}, \\
        \mathcal{C}^-_{\Delta,p}b^\dagger_{X,P}, & \mbox{if $O^\dagger$ is in the future of $X$}.
    \end{cases}
\end{equation}
Here the coefficients $\mathcal C^\pm_{\Delta, p}$ are given in \eqref{normalization1}.

\subsubsection{Direction of modes and notation} 
Near a bulk point, the early-time physics consists of \textit{left moving modes} which are causally connected to the past boundary hyperboloid, while the late-time physics consists of \textit{right moving modes} which are causally connected to the future hyperboloid, with both hyperboloids lying on the right boundary\footnote{The  directionality of modes is defined with respect to positive energy, and works well for negative energy modes as well if we pull an overall minus sign out.}. The notation corresponds to the visual representation of positive frequency modes on the Penrose diagram.

We will utilize the following notation for the left-moving bulk momenta $P$ as a function of boundary momenta $p$, where we consider the positive-frequency mode:
\beq
P  =(p_\mu, P_z)=(-\omega_p, {\bf p},  P_z ), \qquad {\rm s.t.} \qquad P \cdot \delta X =\, p_\mu \delta X^\mu + P_z \, \delta  z, 
\eeq
The early-time modes $a_{X,P} / a^\dagger_{X,P}$ accompany the phases $e^{\pm \ic P \cdot \delta X}$.  Similarly for the right-moving modes, we define the bulk momenta $Q$ in terms of the boundary shooting momenta $q$.
\beq
Q  =(q_\mu, - Q_z )=(-\omega_q, {\bf q},  -Q_z ), \qquad {\rm s.t.} \qquad Q \cdot \delta X =\, q_\mu \delta X^\mu - Q_z \, \delta  z,
\eeq 
such that late-time oscillator modes $b_{X,Q} / b^\dagger_{X,Q}$ accompany the phases $e^{\pm \ic Q \cdot \delta X}$. 

\subsubsection{Factorization to flat-space like amplitudes} 
Let us now consider a four-point correlator with at least one past operator insertion and one future operator insertion. The generic correlator is represented by the following form:
$$\langle \Psi  | \ {\rm Product \, of} \  \  \!O_{x,p,\sigma}{\rm 's}, \  \!O^\dagger_{x,p,\sigma}{\rm 's}, \   O_{y,q,\sigma}{\rm 's} \  O_{y,q,\sigma}^\dagger{\rm 's} \ |  \Psi \rangle$$
To explain this generic form, different operator arrangements can lead to a different Lorentzian correlators with different timefolds. As an example, for the radar correlator, we have the following configuration:
$$
\langle \Psi |  \, O_{x_4, p_4, \sigma_4} \, O_{y_3, q_3, \sigma_3}\, O^\dagger_{x_2, p_2, \sigma_2} \, O^\dagger_{x_1, p_1, \sigma_1} \,  | \Psi \rangle
$$
As another example, we can also consider an otoc lying on two timefolds, with the following operator ordering:
$$
\langle \Psi |  \, O^\dagger_{y_2, q_2, \sigma_2} \, O_{x_4, p_4, \sigma_4} \, O_{y_3, q_3, \sigma_3}\,   O^\dagger_{x_1, p_1, \sigma_1} \,  | \Psi \rangle
$$
Both these correlators describe exterior processes, which can pick up a bulk point if the boundary shooting momenta are directed properly \cite{Caron-Huot:2025she}. Let us understand this case here. Inserting \eqref{eq:dictionarysmeared} into the generic form of the  correlator gives us a factorization formula:
\beq \label{rep2}
\begin{split}
\langle \Psi  | \ & {\rm Product \, of} \  \  \!O_{x,p,\sigma}{\rm 's}, \  \!O^\dagger_{x,p,\sigma}{\rm 's}, \   O_{y,q,\sigma}{\rm 's} \  O_{y,q,\sigma}^\dagger{\rm 's} \ |  \Psi \rangle \\
 & \approx \int \qty[\prod \diff P_i\,\psi_i] \qty[\prod \diff Q_j\,\psi_j] \,  \expval{\ {\rm Product \, of} \  \  \!a_{P}{\rm 's},  \ a^\dagger_{P}{\rm 's},\ b_{Q}{\rm 's}, \ b^\dagger_{Q}{\rm 's}}{0}^{\text{near X}}, \\
 & \approx \int \qty[\prod \diff P_i\,\psi_i] \qty[\prod \diff Q_j\,\psi_j] \, \sqrt{-g} \, (2\pi)^{d+1}\ \delta^{d+1}\qty(\textstyle{\sum \alpha_i P_i + \sum \alpha_j Q_j})\, \ \ic \cM(\{P_i, Q_j\}), \\
\end{split}
\eeq
where $\alpha_k = \pm 1$ take into account the direction of the boundary shooting momenta. Note that the bulk inclusive amplitude has the same operator ordering as that of the boundary out-of-time-ordered correlator. The factor $\ic \cM(\{P_i, Q_j\})$ denotes the inclusive flat space Feynman rule obtained using the Lagrangian governing the bulk interaction.

\subsection{Inclusive observables and the four point correlator}
\label{incobv4pt}

Let us consider the expectation value of a late-time operator insertion $\phi_{p_3}$ over an initial state given by a superposition of different particles: 
\beq
\langle \psi \, | \, \phi_{p_3} \, | \, \psi \rangle, \qquad \text{where} \qquad | \, \psi \rangle = a^\dagger_{p_1} a^\dagger_{p_2} \  | \Omega \rangle + a^\dagger_{p_4} \  | \Omega \rangle
\eeq
Here $ a_{p_3} $ and $a^\dagger_{p_3}$ denote early-time absorption and emission modes respectively. We decompose the field $\phi_{p_3}$ in terms of late-time absorption modes $ b_{p_3}  $ and emission modes  $ b^\dagger_{p_3}$. The expectation value $\langle \psi \, | \, \phi_{p_3} \, | \, \psi \rangle$ contains terms which are expectation values over a particle-number eigenstate.  Additionally we also have the cross-term contributions:
\beq
\langle \Omega \, | \, a_{p_4} \, b_{p_3} \,  a^\dagger_{p_2} \, a^\dagger_{p_1} \, | \, \Omega \rangle  \ + \ \langle \Omega \, | \, a_{p_1} \,  a_{p_2} \, b_{p_3} \,  a^\dagger_{p_4} \, | \, \Omega \rangle  \ + \ {\rm h.c.}
\eeq
Each of these terms with mixed oscillators describe scattering amplitudes with insertions on a Schwinger Keldysh fold.

As another motivation for inclusive observables, let us revisit the two-point Hawking correlator in the context of evaporating black holes \cite{Hawking-particle-creation}. The early time modes here are labelled using $e / e^\dagger$, and the late time modes are labelled using $f / f^\dagger$. Consider the in-state corresponding to $e_\textbf{k} | {\Omega} \rangle = 0$, where $|\Omega\rangle$ denotes the initial vacuum state of the black hole. We are interested in the following in-in correlator involving the late time number operator $N_{\textbf{k}} = f^\dagger_\textbf{k} f_\textbf{k}$:
\beq
\begin{split}
\langle \Omega | \, f^\dagger_\textbf{k} f_\textbf{k} \, | \Omega \rangle &= \sum_{\rm out} \langle \Omega | \, f^\dagger_\textbf{k} \,  |{\rm out} \rangle  \, \langle {\rm out}| \,  f_\textbf{k} \, | \Omega \rangle  \\
& = {B({ k}) \over e^{\beta \omega_{\bf k}} -1}  \ \delta^d(k -k')
\end{split}
\eeq
where $B( k)$ denotes multiplicative greybody factors. On the rhs, we have decomposed the in-in correlator into products of in-out amplitudes taking us from the black hole state to some final-microstate $|{\rm out} \rangle$, which we do not need to measure explicitly to calculate the final answer. Related to this, \cite{Aoude:2024sve} specify Bogoliubov coefficients in terms of generalized in-in amplitudes.

\paragraph{Time-ordered correlators:}
Consider the time-ordered CFT correlator 
$$\langle \Psi | \, O_{y_4, q_4, \sigma_4} \, O_{y_3, q_3, \sigma_3} \, O^\dagger_{x_2, p_2, \sigma_2} \, O^\dagger_{x_1, p_1, \sigma_1} \,  | \Psi \rangle$$ 
As discussed, the conformal correlators factorize to provide inclusive flat space amplitudes in the bulk description. The future timefold allows us to relate the $2\to 2$ time-ordered inclusive amplitude to a time-ordered in-out scattering amplitude as follows.
\beq \label{ininex3}
\begin{split}
\langle \ b_{X,Q_4} \, b_{X,Q_3} \,  a^\dagger_{X,P_2} \, a^\dagger_{X,P_1}  \, \rangle  = \sum_{\rm out} \langle \Omega   |  {\rm out} \rangle \, \langle {\rm out}  | \ b_{X,Q_4} \, b_{X,Q_3} \, a^\dagger_{X,P_2} \, a^\dagger_{X,P_1}  \, \rangle
\end{split}
\eeq
One can perform in-out scattering experiments designed with the in-out S-matrix element on rhs of \eqref{ininex3}. However, this involves the inconvenience of acquiring the final microstate ($|{\rm out} \rangle$) which slightly complicates the experiment. Summing over the final microstate is a convenient way out, which is naturally implemented by considering correlators on the in-in timefold.

\subsubsection{Experiments using in-in timefolds}
When the bulk point $X$ is close to the horizon, where some scattering end-products may end up falling inside it making the experiment difficult. Operator insertions on the in-in timefold circumvent this issue and allows for a larger variety of experiments compared to in-out processes. 

We consider scattering processes with at least one boundary operator insertion on the past boundary hyperboloid $x$ as well as one on the future hyperboloid $y$. We basically take: 
$$b_{X,Q} \, |\Omega \rangle \quad ({\rm late \ time}) \quad \mapsto \quad a_{X,P} \ |\Omega \rangle  \quad ({ \rm early \ time})$$
where the local bulk momenta are related by $P_M = -Q_M$. This action is allowed since in flat-space annihilation operators annihilate the vacuum, i.e. $b_{X,Q} \ | \Omega \rangle = a_{X,Q} \ | \Omega \rangle = 0$. 

Consider the correlator in Fig \ref{fig:extbpcontour} where the above procedure gives us a $3 \to 1$ scattering experiment.  This describes a bulk point in the black hole exterior, as shown in Fig \ref{mainfig00}. The future timefold again introduces a complete sum over future states, which defines the radar experiment.

\section{Boundary features of bulk scattering}
At the level of the four point function, scattering in the bulk has a distinct boundary signature. Consider the Euclidean signature. Using conformal invariance, we can write the four point correlator in terms of an invariant function $\mathcal{G}(z,\bar{z})$ times a kinematical factor.
\beq 
 \langle \Psi| \ O(x_1) \ O(x_3) \ O(x_2) \ O(x_4) \, |\Psi \rangle =
 \frac{1}{x_{24}^{2\Delta_O}\ x_{13}^{2\Delta_{O}}} \mathcal{G}(z,\bar{z})
\eeq
where $x_{ij}$ denotes the spacetime distance between the points $i$ and $j$. Here the positional dependence of the correlator is captured by cross ratios, defined as follows:
\beq \label{cratios}
 \frac{x_{12}^2 \ x_{34}^2}{x_{13}^2 \ x_{24}^2}\equiv z\, \bar{z},\qquad
 \frac{x_{14}^2 \ x_{23}^2}{x_{24}^2 \ x_{13}^2}\equiv(1{-}z)\,(1{-}\bar{z}). 
\eeq
This implies the cross-ratios can be expressed as follows.
\beq 
 z = \frac{z_{12} \ z_{34}}{z_{13} \ z_{24}}, \qquad \bar{z} = \frac{\bar{z}_{12} \ \bar{z}_{34}}{\bar{z}_{13} \ \bar{z}_{24}}, \qquad \quad 1-z = \frac{z_{14} \ z_{23}}{z_{13} \ z_{24}}, \qquad 1-\bar{z} = \frac{\bar{z}_{14} \ \bar{z}_{23}}{\bar{z}_{13} \ \bar{z}_{24}}
\eeq
In the scalar sector, the operators with weights $\Delta_O$ are dual to a scalar field in AdS$_{d+1}$ spacetime via the mass dimension relation (with $R_{\rm AdS} = 1$).
\beq \label{eq5.2}
\Delta=\frac{d}{2}+\nu, \qquad \nu = \sqrt{{d^2 \over 4} + m^2}
\eeq 
In the Lorentzian signature, the general correlator is defined by implementing the $\ic \epsilon$ prescription for the operators appropriately, see \cite{Hartman:2015lfa, Haehl:2017qfl} for a detailed description. 
\subsection{Vacuum scattering experiments}
We discuss the two classes of experiments: the radar experiment and an otoc with two future timefolds.
\subsubsection{Radar correlator}
 In the position space, the radar configuration corresponds to the following correlator with the following arrangement.
\beq
\langle \Omega | \, O(x_2)\, O(y_4)\, O(x_3)\, O(x_1) \, | \Omega \rangle \ = \
\begin{tikzpicture}[scale=0.8, baseline={(0,-0.2)}]
  \draw (0,0.1) -- (0,0) -- (5.5,0) -- (5.5,-0.3) -- (0,-0.3) -- (0,-0.4);
  \node [below] at (1.2,-0.3) {$O(x_2)$};
  \node [above] at (4.5,0) {$O(y_4)$};
  \node [above] at (0.6,0) {$O(x_1)$};
  \node [above] at (2.4,0) {$O(x_3)$};
  \node [right] at (5.5,0.1) {\small $1$};
  \node [right] at (5.5,-0.25) {\small $2$};
  \filldraw (4.5,0) circle (2pt);
  \filldraw (1.5,-0.3) circle (2pt);
  \filldraw (1,0) circle (2pt);
  \filldraw (2,0) circle (2pt);
\end{tikzpicture}
\eeq
Here $O(y_4)$ lies within the future lightcone of the rest of the operators. From an initial Euclidean configuration where all $x_i$ are spacelike separated such that $z = \bar{z}$ and $0<z<1$, we bring the point $y_4$ inside the lightcones. 
\begin{figure}[t!]
    \centering
    \begin{subfigure}{0.45\textwidth}
    \centering
\begin{tikzpicture}[scale=3.5]
    \node[label=below:{$1$}] (X) at (3.55,0) {};
\node[label=below:{$2$}] (X) at (3.85,0) {};
\node[label=below:{$3$}] (X) at (4.15,0) {};
\node[label=below:{$4$}] (X) at (4.5,0) {};
\draw[dashed] (3.55,0) -- (3.05,0.5) (3.55,0) -- (4.05,0.5) (3.85,0) -- (4.35,0.5) (3.85,0) -- (3.35,0.5) (4.15,0) -- (4.65,0.5) (4.15,0) -- (3.65,0.5); 
\draw[->, red, thick] (4.5,0) -- (3.85,0.65);
\end{tikzpicture}
    \caption{}  \label{redpath}
\end{subfigure}
\begin{subfigure}{0.45\textwidth}
    \centering
\begin{tikzpicture}[scale=3.5]
    \node[label=below:{$1$}] (X) at (3.55,0) {};
\node[label=below:{$2$}] (X) at (3.85,0) {};
\node[label=below:{$3$}] (X) at (4.15,0) {};
\node[label=below:{$4$}] (X) at (4.5,0) {};
\draw[dashed] (3.55,0) -- (3.05,0.5) (3.55,0) -- (4.05,0.5) (3.85,0) -- (4.35,0.5) (3.85,0) -- (3.35,0.5); 
\draw[->, red, thick] (4.5,0) -- (3.85,0.65);
\draw[->, red, thick] (4.15,0) -- (3.65,0.5);  
    \draw[->] (5,0.3)--(5.1,0.2);
    \node [thick, right] at (5.1,0.15) {$z$};  
        \draw[->] (5,0.3)--(5.1,0.4);
    \node [thick, right] at (5.1,0.45) {$\bar{z}$};  
\end{tikzpicture}
    \caption{}  \label{redpaths}
\end{subfigure}
    \caption{(a) Going from the Euclidean to the radar configuration. Crossing these light cones takes us to the second sheet and eventually to the bulk point discontinuity for scattering configuration as in Fig \ref{fig:extbpcontour}. (b) Going from the Euclidean to the otoc configuration on two timefolds. Crossing lightcones takes us to the second sheet and from there to the bulk point singularity.}
\end{figure}
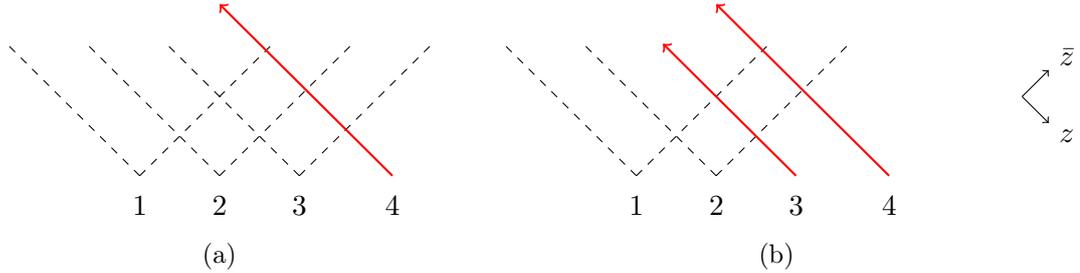
Corresponding to the red path in Fig \ref{redpath}, it can be checked that the $\ic \epsilon$ prescription leads to the following behaviour of the cross ratio $z$ about the branch points at $z =1$ and $z =0$ respectively:
\beq
\begin{split}
 1-z = \frac{z_{14} \ z_{23}}{z_{13} \ z_{24}} \sim \frac{(t_{41} - \ic \epsilon)\ t_{32}}{t_{31} \ (t_{42} + \ic \epsilon)}, \qquad 
 z = \frac{z_{12} \ z_{34}}{z_{13} \ z_{24}} \sim \frac{(t_{43} - \ic \epsilon)\ t_{21}}{t_{31} \ (t_{42} + \ic \epsilon)}
\end{split}
 \eeq
This implies that the cross ratio $z$ undergoes a clockwise $2 \pi$ rotation about the branch cut starting at $z = 1$ and a $2 \pi$ counter-clockwise rotation about the branch cut starting at $z=0$ and lies on the scattering sheet. Additionally from Fig \ref{redpath}, the configuration of $\bar{z}$ is relatively unchanged, i.e. it does not cross any branch cuts.

\subsubsection{Otoc with two-timefolds} 
Apart from the radar experiment, it is also useful to consider another experiment capturing the four point correlator with two future timefolds. Consider the otoc $2\to 2$ experiment described by the following correlator:
\beq \label{otoc with two tfs}
\langle \Omega | \ O(y_3)\ O(x_2)\ O(y_4)\ O(x_1) \ | \Omega \rangle \ = \
\begin{tikzpicture}[scale=0.8, baseline={(0,-0.5)}] 
\draw  (0,0.1)--(0,0) -- (5.5,0) -- (5.5,-0.3) -- (0,-0.3)--(0,-0.6) -- (5.5,-0.6) -- (5.5,-0.9) -- (0,-0.9) --(0, -1);
\node [below] at (1.8,-0.8) {$O(x_2)$};
\node [above] at (4.5,-0) {$O(y_4))$};
\node [above] at (0.6,0) {$O(x_1)$};
\node [below] at (4.5,-0.8) {$O(y_3)$};
\filldraw (4.3,0) circle (2pt); \filldraw (4.6,-0.9) circle (2pt); \filldraw (1,0) circle (2pt); \filldraw (1.3,-0.3) circle (2pt);
\end{tikzpicture}
\eeq
We can start from the Euclidean configuration where $0<z<1$ in the initial configuration as shown in Fig \ref{redpaths}. The red paths in Fig \ref{redpaths} imply that the cross ratio $z$ takes the following values for the final configuration:
\beq
z \sim {t_{21} \ t_{43} \over (t_{31} - \ic \epsilon_{13}) \ (t_{42} + \ic \epsilon_{42})}
\eeq
Note from the contour that $\epsilon_{13}> \epsilon_{42}>0$, which implies $z$ goes counter-clockwise about $z=0$. Similarly about $z=1$, the cross-ratio moves as:
\beq
1-z \sim {(t_{41} - \ic \epsilon) \ (t_{32} -\ic \epsilon)  \over (t_{31} - \ic \epsilon_{13}) \ (t_{42} + \ic \epsilon_{42})}
\eeq
Again as in the radar configuration, the behaviour of the cross-ratios implies that the cross ratio $z$ goes $2\pi$ clockwise around the branch cut at $z=1$, while it goes $2 \pi$ anti-clockwise about $z=0$. Consequently we have scattering configuration.

\subsection{Reading off the bulk momenta}
The path in Fig \ref{redpath} leads to bulk scattering. In this limit, the amplitude goes as power law in terms of the difference of the cross ratios $z$ and $\bar z$:
\beq \label{div}
\mathcal{G}(z,\bar{z}) \sim {1 \over (z-\bar{z})^n}
\eeq
where $n>0$ \cite{Gary:2009ae}. As $z \to \bar z$, we encounter the bulk point singularity. Here scattering over the vacuum indicates $n = d-3$ \cite{Maldacena:2015iua}. For $d=2$, we find $\mathcal{G}_{d=2}(z, \bar{z})\to 0$ as we approach the discontinuity while for $d=4$ there is a genuine pole. See Appendix \ref{sec:morebp} for further explanation why the bulk point singularity arises from D-function and Mellin space representations.

Here we use the expression \eqref{div} for the bulk amplitude and use it to read the bulk momenta about the bulk point in terms of a boundary measurement. Let us start with the Schwinger representation of the discontinuity in \eqref{div}:
\beq
{1 \over (\delta z)^n} = {{ e^{-\ic \pi n \over 2}} \over \Gamma(n)} \int_0^\infty {{\rm d}s \over s} s^n\exp  \lc {\ic s \delta z }  \rc
\eeq
Here $\delta z = z(x_j) - \bar{z}(x_j)$. 

\subsubsection{Wavepackets}
We introduce wavepackets by integrating the discontinuity against smeared Fourier transforms. We integrate over $\delta x_j$ in the vicinity of the bulk point. In the plane wave limit, the bulk amplitude is found by integrating over the elevator region about the bulk point \cite{Maldacena:2015iua, Chandorkar:2021viw}, which gives us the following form:
\beq
\mathcal{M}(p_i) = {{ e^{-\ic \pi n \over 2}} \over \Gamma(n)} \int_0^\infty {{\rm d}s \over s} s^n  \prod_{j=1}^4\int {\rm d}^d (\delta x_j) \ e^{\ic p^j_\mu \, \delta x_j^\mu + \ic s \,  \delta z}  
\eeq
This is a highly oscillatory integral. Here the Fourier transforms originate since the locus lies on the boundary hyperboloids which depends on the shooting angle and the bulk point as $x = x(p,X)$. 

The integral is dominated by the saddle points of the phase function:
\beq \label{phase:Phi}
\Phi = \sum_j p_j \cdot \delta x_j + s \, (\delta z).
\eeq
Note that the leading phase is independent of $n$. We vary the phase above with respect to the boundary insertion $x_j^\mu$. This gives us the stationarity condition using which the boundary momenta can be written as:
\beq
\label{eq:saddle_p}
p_{j\mu} + s \frac{\partial (\delta z)}{\partial (\delta x_j^\mu)} = 0.
\eeq
The above equation identifies the boundary momentum $p_j$ with the partial derivative of the difference in cross-ratios $\delta z$ upto the Schwinger parameter $s$.

\subsubsection{Bulk momenta}
We clarify our notation here: the capital letters and Latin indices denote bulk quantities, while small letters and Greek indices denote boundary quantities. To read off the bulk momentum $P_{j M}$ associated with leg $j$ at the bulk point $X$ we utilize the Hamilton-Jacobi formulation of geodesics of \cite{Caron-Huot:2025she}. These geodesics arise in the geometric optics description and connect the bulk point to the boundary hyperboloids.

We recall that in the high frequency WKB approximation, the phase acquired by propagating from the boundary point $x_j$ to the bulk point $X$ takes the following form:
\beq
S_j(p_j, X) =  p_{j\mu} \ x_j^\mu(p_j, X) +  S_j(x_j, X), \qquad {\rm high  \ energy} 
\eeq
The above equation is a canonical transformation of the massless geodesic action $S_j(x_j, X)$. The on-shell contribution of leg $j$ to the phase is $S_j(p_j, X) = p_{j\mu} \ x_j^\mu$ for a massless geodesic since the on-shell value of $S_j(x_j, X)$ is zero.  The bulk momentum is the variation of the phase $S_j(p_j, X)$ with respect to the bulk coordinate $X^M$:
\beq
P_{j M} = \frac{\partial S_j(p_j, X)}{\partial X^M} = \frac{\partial}{\partial X^M} \lc p_j^\mu \ x_{j\mu}(X) \rc = p_j^\mu \frac{\partial x_{j\mu}}{\partial X^M}.
\eeq
We can now utilize the boundary momenta using the saddle point equation which determines the boundary momentum as in \eqref{eq:saddle_p}. Substituting this into the expression for bulk momentum, we obtain the following expression for the bulk momenta:
\beq
P_{j M} = -s \ \frac{\partial (\delta z)}{\partial ( \delta x_j^\mu)} \ \frac{\partial x_j^\mu}{\partial X^M}.
\eeq
We can simplify this equation further by trading the Schwinger parameter for a boundary vector, which could be one of the boundary momenta or any other vector of choice. Here we can project $s$ using \eqref{eq:saddle_p} along an arbitrary boundary vector $\xi^\mu$:
\beq
s = - \frac{p_{1\mu} \xi^\mu}{\frac{\partial (\delta z)}{\partial ( \delta x_1^\nu )} \xi^\nu }.
\eeq
Here we have utilized the variation in $(p_1  \cdot \delta x_1)$. Substituting this back into the expression for $P_{j M}$, we find that the bulk momenta are given by:
\beq \label{bulk mom formula}
P_{jM} = \lc \frac{p_{1\mu} \xi^\mu}{\frac{\partial (\delta z)}{\partial ( \delta x_1^\nu )} \xi^\nu }\rc  \frac{\partial (\delta z)}{\partial ( \delta x_j^\mu)} \ \frac{\partial x_j^\mu}{\partial X^M}.
\eeq
This formula cleanly expresses the local bulk momentum $P_M$ in terms of the boundary kinematic data $p_\mu$ about a sharp bulk point singularity. 

\subsubsection{General geometries} 
The formula \eqref{bulk mom formula} readily admits a generalization to general geometries such as the black hole exterior, where the validity of the high-frequency WKB solution allows us to the easily generalize this. 

Corresponding to the bulk point singularity in a general geometry \cite{Caron-Huot:2025hmk}, the factorization into the flat space amplitude leads to a universal form which is independent of the bulk location and the geometry. This implies that we have a bulk point singularity of the form \eqref{div}. 

There are three ingredients here, firstly the WKB solution is matched to the near-boundary solution where it becomes empty AdS. The next part is the bulk evolution using WKB. The covariant form of the on-shell action illustrates this cleanly:
\beq
S_j(p_j, X) = p_{j\mu} \ x_j^\mu (p_j, X), \qquad {\rm high \ energy}
\eeq
Here the massless limit implies that with any choice of background geometry we obtain a universal boundary contribution. Finally near the bulk point $X$, the phase looks as
\beq
   \exp(\ic S_j(p_j; X + \updelta X)) = \exp(\ic S_j(p_j; X) + \ic P_{j,M} \updelta X^M + \mathcal{O}(\updelta X^2)).
\eeq
Since all steps in the construction are unaltered, we obtain the same expression for $P_{j M}$ where the bulk momenta are given by the formula \eqref{bulk mom formula}.


\subsection{Scattering in AdS-Rindler spacetime}
It will be useful to chart out a thorough discussion in three-dimensional AdS-Rindler space prior to undertaking a detailed study of black holes. Our goal here is to mainly understand how in the cross ratio space we approach the bulk point kinematics. 

We start with Rindler metric, study the hyperboloids and then the extension to the future interior region corresponding to an infalling geodesic. We then study a path with takes the exterior bulk point to an interior configuration as in Fig \ref{fig:crr}.

Consider the following representation of Rindler spacetime in three dimensions, which takes the following form:
\beq
{\rm d}s^2 = {1 \over z^2} \lc -h(z){\rm d}T^2 + {{\rm d}z^2 \over h(z)} + {\rm d}x^2 \rc
\eeq
Here $h(z) = 1-z^2$ and the inverse temperature is given by $\beta = 2 \pi $. The coordinates resemble the BTZ black hole metric, with the exterior region given by $z<1$.

\paragraph{Boundary hyperboloids:} Let us firstly discuss the boundary imprint of a lightcone from the a point $X$, i.e. the boundary hyperboloids in Rindler spacetime.The boundary hyperboloids are lightlike curves from a bulk point $X = (T, { X}, z)$ to boundary locus $y = (t, { y})$, which can be obtained by integrating the geodesic equations:
\beq 
    t - T = \pm \omega \int^{z}_0 \frac{{\rm d}z}{h(z) \sqrt{\omega^2  - { p}^2 h(z)}}; \qquad 
{y} - {X} =  { p} \int^{z}_0 \frac{{\rm d}z}{\sqrt{\omega^2  - { p}^2 h(z)}}
\eeq
The exact expressions for the hyperboloids in the exterior region $z<1$ are given by:
\beq \label{Rindler}
    t = {\beta \over 2 \pi} \arccoth\lc \frac{\beta}{2 \pi z} \frac{\sqrt{\omega^2  - { p}^2 h(z)}}{\pm \omega}\rc + T, \qquad { y} = {\beta \over 2 \pi} \arccoth \lc \frac{\beta}{2 \pi z} \frac{\sqrt{\omega^2  - { p}^2 h(z)}}{{ p}}\rc + { X},
\eeq
where we have explicitly kept $\beta$ in our expressions to match with later generalizations.

\subsubsection{Infalling geodesic into the future region}
 Near the horizon, the temporal component of the geodesic equation simplifies in terms of the expression for the Tortoise coordinate. More precisely, this happens since we have the following limit:
\beq
\omega \int \frac{{\rm d}z}{h(z) \sqrt{\omega^2  - { p}^2 h(z)}} \approx \pm \int \frac{{\rm d}z}{h(z)} \equiv z_*(z) ; \qquad {\rm as} \qquad h(z) \to 0.
\eeq
Near horizon, since the geodesic equation simplifies in terms of the above Tortoise coordinate, we can use it to venture behind it. For an infalling geodesic, it is convenient to utilize the lightcone coordinates defined as follows:
\beq
U = -e^{{2 \pi \over \beta} (-T +z_*)}; \qquad V = e^{{2 \pi \over \beta} (T +z_*)};
\eeq
These lightcone coordinates motivate us to keep the black hole picture of Fig \ref{fig:rindler} in mind for our present discussion since the near-horizon geometry of Schwarzchild solution is two-dimensional Rindler. The form of the lightcone coordinates here implies a constant null coordinate $v$ as expected for an infalling geodesic. 

In terms of the original Rindler coordinates for a future-directed infalling geodesic, near the horizon, we require the following analytic continuations which corresponds to a constant lightcone coordinate $v$:
\beq
 T|_{{\rm Re} z<1 }  \to T|_{{\rm Re} z>1 } - {\ic \beta \over 4}; \qquad z_*|_{{\rm Re} z<1 } \to z_*|_{{\rm Re} z>1 } + {\ic \beta \over 4}
\eeq
Within the context of our experiment, these correspond to the early time insertions which lead to infalling geodesics. We also have outgoing rays with constant $U$ connected to the same bulk point which take us out from the interior to the other exterior. For these past directed outgoing geodesics, the above continuation requires that the future insertions upon crossing the horizon are connected to the boundary at:
\beq
t \to -t - {\ic \beta \over 2}; \qquad T|_{{\rm Re} z>1 }  \to T|_{{\rm Re} z<1 } - {\ic \beta \over 4}; \qquad z_*|_{{\rm Re} z>1 } \to z_*|_{{\rm Re} z<1 } - {\ic \beta \over 4}
\eeq
This corresponds to shifting the late-time insertions to the other exterior Rindler wedge. Within our context, these will correspond to geodesics that have been smeared and pushed from the future region to the left CFT. For more details see Appendix \ref{2drindler}.

\subsubsection{Boundary picture}

We analyze the details of our proposal for two dimensional CFT from the boundary side.
 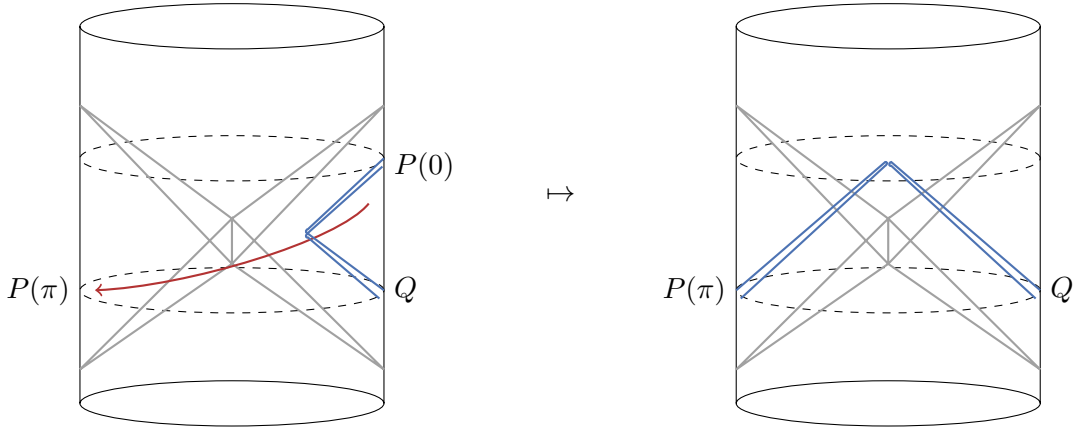
\begin{figure}[t!]
    \centering
\begin{tikzpicture}[scale = 1.0, xscale=2.0]
\draw (-1.0,2.25) -- (-1.0,-2.75)  (1,-2.75) arc (0:360:1 and 0.3) -- (1,2.25) arc (0:360:1 and 0.3);
\draw[dashed] (1,-1.25) arc (0:360:1 and 0.3) (1,0.5) arc (0:360:1 and 0.3);
\draw[mgray,thick]  (1,1.2) -- (0,-0.3) (0,-0.9) -- (1,1.2)  (-1,1.2) -- (0,-0.3) (0,-0.9) -- (-1,1.2) (1,-2.3) -- (0,-0.3) (0,-0.9) -- (1,-2.3) (-1,-2.3) -- (0,-0.3)--(0,-0.9) -- (-1,-2.3);
\draw[mred,thick, ->] (.9,-0.1) ..controls (0.7,-0.6) and (-0.3,-1.2).. (-0.9,-1.25);
\draw[mblue, thick] (1,-1.25) --(0.5,-0.5) to[out=200,in=-150] (0.5,-0.55) --  (0.97,-1.36) (1,0.5) --(0.5,-0.44) to[out=200,in=-150] (0.5,-0.49) --  (0.99,0.39);
\node [right] at (1,-1.25) {$Q$}; 
\node [right] at (1,0.39) {$P(0)$}; 
\node [left] at (-1,-1.25) {$P(\pi)$}; 
\node [right] at (2,0) {$ \mapsto \qquad $}; 
\end{tikzpicture} 
\begin{tikzpicture}[scale = 1.0, xscale=2.0]
\draw (-1.0,2.25) -- (-1.0,-2.75)  (1,-2.75) arc (0:360:1 and 0.3) -- (1,2.25) arc (0:360:1 and 0.3);
\draw[dashed] (1,-1.25) arc (0:360:1 and 0.3) (1,0.5) arc (0:360:1 and 0.3);
\draw[mgray,thick]  (1,1.2) -- (0,-0.3) (0,-0.9) -- (1,1.2)  (-1,1.2) -- (0,-0.3) (0,-0.9) -- (-1,1.2) (1,-2.3) -- (0,-0.3) (0,-0.9) -- (1,-2.3) (-1,-2.3) -- (0,-0.3)--(0,-0.9) -- (-1,-2.3);
\draw[mblue, thick] (1,-1.25) --(0.02,0.45) to[out=200,in=-150] (0.02,0.4) --  (0.97,-1.36) (-1,-1.25) --(-0.02,0.45) to[out=-350,in=50] (-0.02,0.4) --  (-0.97,-1.36) ;
\node [right] at (1,-1.25) {$Q$}; 
\node [left] at (-1,-1.25) {$P(\pi)$}; 
\end{tikzpicture}  
 \caption{Analytic continuation along the complexified paths $P^\pm(\theta)$ (shown in red) takes us from the bulk point within the Rindler wedge to the future wedge by analytically continuing to the complex coordinates. This takes observers from support on future time in right Rindler wedge to the left Rindler wedge.}  \label{fig:crr}
\end{figure}  
We motivate our construction by studying the example of two dimensional boundary dual to the Rindler wedge of AdS$_3$. As in Fig \ref{fig:crr}, we chart a path along which the boundary processes can be analytically continued which takes us from a bulk point in the exterior region to a bulk point in the interior region.

\paragraph{Complex paths $P^\pm(\theta)$ on boundary:} Motivated by the bulk analytic continuations, let us now design a complex path that takes us to the interior while staying close to the horizon. Keeping real momenta $p_\mu$, we choose bulk points along a complex path from Re $z>1$ to Re $z<1$ which takes into account the branch cut at $z = 1$. In this analysis we require the bulk points to remain close enough to $z = 1$ so that the two dimensional Rindler picture continues to remain valid. This path takes the bulk point into the future region on the complexified geometry. The real geometry are special instances of our one-parameter analysis.

Very close to the horizon, we take the bulk point to move along the bulk time:
\beq
T \to T - {\ic \theta}; \qquad z_* \to z_* + {\ic \theta}
\eeq
while keeping the transverse ${ X}$ real along the path from $\theta \in \lc 0, {\beta \over 4} \rc$. In the $UV$ plane, this corresponds to a near-horizon rotation $ U \mapsto U e^{2 \ic \theta}, \  V \mapsto V$ to the interior,  where initially $U<0, \ V>0$ in the exterior region. From the interior to the exterior, over the parameter range $\theta \in \lc {\beta \over 4}, {\beta \over 2} \rc$, we follow the path given by:
\beq
T \to T - {\ic \theta}; \qquad z_* \to z_* - {\ic \theta}
\eeq
This corresponds to a near-horizon rotation  from interior $ U_I \mapsto U_I , \  V_I \mapsto V_I e^{2 \ic \theta}$ to the  other exterior region, where initially $U_I>0, \ V_I>0$ in the interior region. 

\subsubsection{Analytically continuing a scattering process}
Note that at values given by $\theta = {\beta \over 4},  {\beta \over 2}$, the path describes a scattering process in the real spacetime where for $\theta = {\beta \over 2}$ it localizes to the interior.
Under this action the boundary temporal coordinates corresponding to the shooting points connected to the above bulk point get transformed to the following values:
\beq
t_f \to -t_f - {\ic \theta}; \qquad t_i \to t_i.
\eeq
Here $\theta \in \lc 0, {\beta \over 2} \rc$, and the label $t_f$ denotes late-time insertions while $t_i$ denotes early-time insertions respectively. In other words, we preserve the initial condition of early-time particles while change the final time of the particles along $\theta$. This is also a complexified path on the boundary, where similar to the bulk description, we find that at $\theta = {\beta \over 2}$, the insertions are continued to the left boundary and we again have a geometric interpretation on the real spacetime. 

Finally for late time absorption modes, the above operations define the path $P^+(\theta)$, we also obtain the path $P^-(\theta)$ by taking the other route, where we continue the times to
\beq
t_f \to -t_f + {\ic \theta}; \qquad t_i \to t_i.
\eeq
This is useful for continuing late time emission modes.

\paragraph{Boundary conformal transformation:} It is convenient to work with the boundary lightcone coordinates $u=t_i -{ x}_i$ and $v=t_i +{ x}_i$. Note that these characterize different physics as compared to the bulk lightcone coordinates $U, \ V$ which involve the radial direction. 
Using these coordinates we utilize a different set of boundary Rindler wedge coordinates which are defined in terms of the following functions.
\beq
\rho_i = \ep^{-u_i}, \qquad \bar{\rho}_i = \ep^{v_i}
\eeq
In these coordinates, the final particles have coordinates given by $\rho_{\rm fin} \mapsto \rho_{\rm fin} \, \ep^{\ic \theta}, \ \ \bar \rho_{\rm fin} \mapsto \bar \rho_{\rm fin} \, \ep^{-\ic \theta}$. We can now use the conformal transformation to the $z, \bar z$ plane which takes the following form:
\beq
\rho(z) = \frac{z}{\left(\sqrt{1-z}+1\right)^2}
\eeq
and with a similar map for the complex conjugate $\bar \rho$. Using this we obtain, $z = {4 e^u\over (1 + e^u)^2}=\text{sech}^2\left(\frac{u}{2}\right)$, which represents the AdS Rindler metric on the complex plane. See also the interesting recent work \cite{Kundu:2025jsm} which shows how $S^1 \times \,$time has a much more richer structure of bulk point singularities.

\subsubsection{Radar experiment from boundary}
Utilizing the Rindler wedge description for the example of the radar correlator, the paths $P^\pm(\theta)$ systematically continues the final time insertion to the left wedge boundary along a smooth path in the cross-ratio space. In the cross ratio space, we go along a path that takes the cross-ratios at the bulk point discontinuity for a bulk point inside the Rindler wedge to another topologically equivalent discontinuity that is in the future region. 

It is straightforward to check about special values of $x_i(p_i, X)$ and $y_i(q_i, X)$, that the path taking the starting point $y_i(q_i, X) \mapsto Ly_i(q_i, X)$, with the latter being the ending point takes us from the discontinuity in the Rindler exterior to the discontinuity in the Rindler interior. It can also be seen that the magnitude of the momenta $q$ going out as a wavepacket towards the boundary determines how much to the left of $z =0$ we approach the bulk point discontinuity. 
\subsection{Holographic cameras behind Rindler horizon} 
\begin{figure}[t!]
\centering
\includegraphics[width=0.3\linewidth]{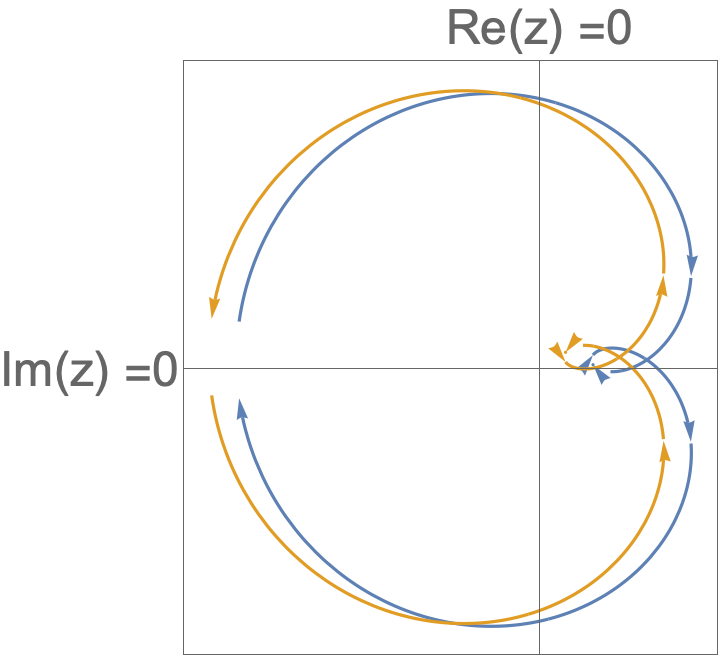} 
    \caption{The path of the cross ratios for the holographic cameras configuration over the complexified red paths $P^\pm(\theta)$ starting from $P(0)$, where we have provided a slight offset in momenta to separate the paths and to clearly see how they land across the bulk point singularity and have kept non-zero $\ic \epsilon$'s. The branch cut starts from $z=0$ and goes to $z = -\infty$. The bulk point singularity is approached at the endpoints $P(0)$ and $P^\pm(\pi)$, where the arrow denotes its approach to $P^\pm(\pi)$.} \label{fig:crrr}
\end{figure}
We want to study the behaviour of cross ratios for insertions which have a shooting momenta. The bulk point $X$ and boundary shooting momenta $p$ give us a locus on the boundary $x(p, X)$, using which we determine the evolution of the cross ratios. 

A particularly simple kinematic limit arises in the coincidence limit, which are the holographic cameras \cite{Caron-Huot:2022lff}. In this limit, the two early time insertions are separated by a small amount $\delta x$ and the two late time insertions are separated by $\delta y$ respectively as in \eqref{otoc with two tfs}. This configuration is on the second sheet as in Fig \ref{redpaths}, where acting with Fourier transforms gives us the bulk scattering. This case has particularly simple kinematics since both the past and the future insertions have equal but opposite momenta which are located in the vicinity of the point $x(p, X) \pm {\delta x /2}$ and $y(q, X) \pm {\delta y /2}$, which allows us to conveniently keep track of the kinematics. 

To understand this case, we can utilize the configuration in the Regge limit where the cross ratios lie in the second sheet. In this limit the cross ratios become $z, \bar{z} \to 0$, and the correlator can be expressed in terms of a sum over conformal Regge trajectories \cite{Costa:2012cb}. The function $\mathcal{G}(z,\bar{z})$ takes the following form in the conformal Regge limit.
\beq \label{eq5.3}
\mathcal{G}(z,\bar{z})-1\approx 2\pi \ic  \int_{0}^{\infty} \frac{d\nu}{2\pi} \, \rho(\nu) \,  \alpha(\nu) \lc z \bar{z} \rc^{\frac{1-J(\nu)}{2}} \, \mathcal{P}_{\frac{2-d}{2}+\ic\nu} \lc \frac{z+\bar{z}}{2\sqrt{z\bar{z}}} \rc
 \eeq
 Note that the integral over the the parameter $\nu$ is given using \eqref{eq5.2}. For a detailed explanation of the various symbols on the right hand side of \eqref{eq5.3} we refer the reader to \cite{Costa:2012cb, Caron-Huot:2022lff}.
 Here $\rho(\nu)$ is the measure, the symbol $\mathcal{P}_J(\eta)$ is the Gegenbauer polynomial for integer $J$ and is given by the an hypergeometric function. The quantity $\alpha(\nu)$ can be derived by studying the Regge behaviour of the corresponding conformal partial wave expansion.

Since we are in the coincidence limit, it suffices to take two Fourier transforms which converts the formula given in \eqref{eq5.3} into its momenta space counterpart \cite{Caron-Huot:2022lff}. This gives them to the following form.
\beq
G(x,p; y,q)-1\approx -\ic  \int_{0}^{\infty} \frac{d\nu}{2\pi} \, \rho(\nu) \,  \tilde{\alpha}(\nu) \lc \abs{p} \abs{q} \rc^{J(\nu)-1} \, \mathcal{P}_{\frac{2-d}{2}+\ic \nu} \lc \hat p . \, \mathcal{I}_{x-y} . \hat q  \rc e^{-{\nu^2 \over 2}\sigma_0^2}
\eeq
This above description is a function of initial coordinates$x,p$ and final coordinate $y,q$, and takes the form of an impact parameter amplitude given in \cite{Cornalba:2006xm, Meltzer:2017rtf, Kulaxizi:2017ixa}. 
\subsubsection{The path in cross-ratio-space} 
We now utilize the path in Fig \ref{fig:crr}, which takes us from the Rindler wedge experiment to the interior experiment. Our starting point in the Figure is the bulk point configuration of holographic cameras where all points are in the same Rindler wedge. This is a well defined experiment in the coincidence limit as given in \cite{Caron-Huot:2022lff}. Our goal in this example is to construct a path in the cross ratio space parametrized by $\theta$ which takes us from the holographic camera setup to a correlator that probes an intersection point in the future Rindler wedge.

Let us write $P^\pm(\theta) = P(\theta)$ here on where it is understood that we are going in opposite directions corresponding to late time emission and absorption operators. Starting with an initial configuration of spacelike operators, crossing lightcones systematically takes us to the second sheet, where we have $z,\bar{z} \to 0$ since we employ the coincidence limit $\delta x, \delta y \to 0$. The Fourier transforms takes the correlator to the bulk point discontinuity at $(P(0), Q)$. Taking $\theta$ from zero to $\pi$ corresponds to taking the points $(P(0), Q)$ on the same Rindler wedge along the red path $P(\theta)$ in Figure \ref{fig:crr} to a configuration $(P(\pi), Q)$ supported on both the wedge and its complement.  Increasing $\theta$, we move around in the cross ratio space, and land up again on the bulk point discontinuity at $(P(\pi), Q)$. They hit the discontinuity now from the opposite sides, thereby preserving the topological configuration. 

In Figure \ref{fig:crrr}, with a slight positive offset to separate the two paths, we show the motion of the two cross ratios along the paths $P^\pm(\theta)$. The bulk point at $(P^\pm(\pi), Q)$ is now outside the original wedge and lies in the future wedge (see Figure \ref{fig:crr}). The bulk point limit is obtained only when the support of the otoc is on the boundary hyperboloids in Figure \ref{fig:crr} such that we obtain the bulk point discontinuity in the cross ratio space.  

 \subsection{Finite-temperature 2d CFT and planar BTZ interior}
 \label{bcft2d}

We discuss the holographic cameras and the radar experiment in the planar BTZ exterior and then take them to the interior. These calculations are enabled due to the conformal transformation from boundary to the plane.

\subsubsection{Holographic cameras}
 We will work with a simple configuration in two dimensional CFT where the transverse direction is spatially infinite. The corresponding bulk dual is the planar BTZ black hole. 
 
 Within this description, we can write thermal correlators in terms of vacuum correlators due to the cylinder to plane map. The exterior discussion here is on similar lines to the single sided exterior otoc in \cite{Caron-Huot:2022lff}. We use the conformal transformation to the plane
 \beq  \label{planbtz}
z_i = \exp \lc \frac{2\pi}{\beta} ({x}_i - t_i)\rc, \qquad \bar{z}_i = \exp \lc \frac{2\pi}{\beta} ({ x}_i + t_i)\rc
 \eeq
 We are particularly interested in the Regge limit, we can utilize formulas from conformal Regge theory \cite{Costa:2012cb}. Specifically we consider the graviton trajectory which contains the $J =2$ resonance. 
 
 For simplicity, we consider the case where all external operator insertions have dimensions $\Delta_{O_i} =1$. We will work in the strong coupling regime where the Regge trajectory $J(\nu)$ stays close to $2$ for high momenta $\nu$ \cite{Costa:2012cb}, where $\nu$ is related to the dimension of the exchanged particle as $\Delta = 1 + \ic\nu$.  The objective here is to evaluate \eqref{eq5.3} for the coincidence limit using the graviton Regge trajectory parametrized below.
  \beq \label{gravRT}
\alpha_{\rm GR}(\nu) = -\frac{G_N \, \pi^2}{2}\frac{(1+\nu^2)}{\cosh^2\lc \frac{\pi \nu}{2}\rc}, \qquad \rho(\nu) = \frac{1}{2}. 
 \eeq
For simplicity, let us set the remaining right CFT locations as follows.
 \beq
x_{1,3} = (0,0) + \frac{\delta x}{2} \lc \cosh \phi_{1} , \sinh \phi_{1} \rc, \qquad  y_{2,4} = ({ y}, t) + \frac{\delta y}{2} \lc \cosh \phi_{2} , \sinh \phi_{2} \rc.
 \eeq
Plugging in the cross ratios using \eqref{cratios} and parametrizing momenta as $p^{\mu} = \abs{p}\lc \cosh \phi_i, \sinh \phi_i \rc$, we Fourier transform the correlator from the position space to the momentum space. Then we arrive at the amplitude given by,
\beq \label{gop}
G(0, p; y, q)\approx -\ic  \int_{-\infty}^{\infty} \frac{d\nu}{2\pi}
  \bar{\alpha}(\nu) \lc \ \abs{p} \, \abs{q}  \ \mathcal{I}(t, { y})\rc e^{\ic\nu (\phi_1 + \phi_2 + \Delta \phi)} e^{-\frac{\nu^2 \sigma_0^2}{2}}
 \eeq
This describes the impact-parameter amplitude in the exterior.

\subsubsection{Interior cameras}
We now study the case when the bulk event takes place in the interior. This is the case where we take $t \to -t \pm {\ic \beta \over 2}$ in the above expressions for the respective emission-absorption operators at late time, which shifts the points $y_i$ from the right boundary to the left boundary. Using this, the intensity $\mathcal{I}(t, { y})$ and $\Delta \phi$ are given by
 \beq \label{Inphi}
\begin{split}
    \mathcal{I}(t, { y}) =\frac{\beta^2}{\pi^2} \cosh  \frac{\pi}{\beta}\lc t + {y} \rc \cosh  \frac{\pi}{\beta}\lc t - {y} \rc, \quad 
    \Delta \phi  = \log \frac{\cosh  \frac{\pi}{\beta}\lc t - {y} \rc}{\cosh  \frac{\pi}{\beta}\lc t + {y} \rc}
\end{split}
 \eeq
 Here we have utilized $\bar{\alpha}(\nu)$ and $\sigma_0$ for the two dimensional case:
 \beq
\bar{\alpha}(\nu) = \frac{4\pi G_N}{(\nu^2 +1)} \qquad \sigma_0^2 = \frac{1}{\sigma_x^2\abs{p}^2} + \frac{1}{\sigma_y^2\abs{q}^2}
 \eeq
 Plugging the expressions for $\mathcal{I}(t, { y})$ and $\Delta \phi$ into \eqref{gop}, we obtain a simple integral expression for the four point correlator. It can be checked numerically that we obtain a sharply peaked signal in time and transverse positions of the boundary detectors as we vary the shooting angles. The sharp features are a smoking gun of a bulk point as was argued in \cite{Caron-Huot:2022lff}.

Let us briefly discuss a couple of important features following from expressions \eqref{Inphi}.  For large time, the function $\mathcal{I}(t, { y})$ grows exponentially as a function of time $\mathcal{I}(t) \sim \exp \frac{2\pi t}{\beta}$ as we asymptote to a large time. This is reflected in the overall magnitude of the signal, and is consistent with previously truncating the signal at a large cutoff time time.  

Finally, in BTZ black hole we expect the bulk point to hit the black hole singularity at $t \to 0$. To obtain this, we need $\phi_x = -\phi_y$ in order to capture the black hole singularity independently of the transverse location. We find from the formulae that the strength of the signal becomes a finite small number upon reaching the black hole singularity. This is unique to three dimensions and intensity of the signal generically disappears in higher dimensions. In \S \ref{sec:bhsing}, we will understand these observations in a better way.

\subsubsection{Radar experiment} 
Let us now move onto the radar experiment on the same background. We will now use the map from the boundary coordinates defined on the cylinder $S_1 \times R$ to the two-dimensional complex plane. 

We now evaluate the cross ratios in terms of these coordinates on the complex plane. The analysis in the previous subsection goes through unchanged: it can be checked that the cross ratios kinematics goes to the scattering sheet. 

To pick up the bulk point, we use sharply directed wavepackets towards a bulk point $X$. We shoot into the bulk using the boundary momentum $p_i = (\omega_i, {\rm p}_i)$ of the four boundary operators.  To obtain the boundary signature, we impose the momentum conservation equations:
\beq
\sum_{i=1}^4 \omega_i = \sum_{i=1}^4 {\rm p}_i =  0
\eeq
We shoot these operators by placing them on a boundary locus corresponding to the bulk point $X = (T,{\rm X},z)$ (we refer the reader to \S \ref{hypeintpt} for details):
\beq \label{eq:planarbtzext}
   t = z_\mathrm{h} \arccoth\lc \frac{z_\mathrm{h}}{z} \frac{\sqrt{\omega^2  - {\rm p}^2 f(z)}}{\pm \omega}\rc + T, \qquad y = z_\mathrm{h} \arccoth \lc \frac{z_\mathrm{h}}{z} \frac{\sqrt{\omega^2  - {\rm p}^2 f(z)}}{{\rm p}}\rc + {\rm X},
\eeq
The sum over all ingoing momenta at the bulk point is zero. To obtain the correct kinematics at the bulk point, we also impose the radial bulk momentum conservation $\sum_{i=1}^4 P_z^i =0$ on the boundary momenta $p_i$. Note that this condition is valid only at the bulk point $X$. 

We find that the boundary correlator displays a scattering signature as in \eqref{div} where the cross ratios approach each other about the branch cut along the real negative-$z$ axis starting from $z=0$.
In the forward scattering case the cross ratios approach each other, i.e. $z \to \bar{z}$ across the branch cut at a slightly negative value of $z$. 

To go to the interior, we again place the operators on a boundary locus corresponding to the bulk point $X = (T,{\rm X},z)$:
\beq \label{eq:planarbtzint}
    t = z_\mathrm{h} \, {\rm arctanh}\lc \frac{z_\mathrm{h}}{z} \frac{\sqrt{\omega^2  - {\rm p}^2 f(z)}}{\pm \omega}\rc + T, \qquad y = z_\mathrm{h} \arccoth \lc \frac{z_\mathrm{h}}{z} \frac{\sqrt{\omega^2  - {\rm p}^2 f(z)}}{{\rm p}}\rc + {\rm X},
\eeq
Here again we impose bulk momentum conservation at $X$. This ensures a scattering signature as in \eqref{div} where the cross ratios approach each other about the branch cut along the real negative-$z$ axis starting from $z=0$.

From the boundary side, we can use conformal Regge theory and work out explicit formulae similar to the case of holographic cameras in the Regge limit. In \S \ref{sec:lsnbp} we will perform a detailed bulk WKB analysis of the correlator and obtain the flat space factorization.  In particular, it is easier to calculate boundary quantities at a finite temperature in two dimensions as compared to finite temperature CFT in higher dimensions due to the map we discussed. 

\section{Local scattering near an interior bulk point}
\label{sec:lsnbp}
We analyze local physics in the vicinity of an interior bulk point. We perform analytic continuations of experiments about a bulk point in black hole exterior to the interior. 

\subsection{Kruskal extension of exterior}

 \begin{figure}[t!]
    \centering
    \begin{tikzpicture}[scale=3.5] 
    \draw [->,line width=0.2mm,domain=-1:1] plot ({\x+ 1.3}, {1*\x});
    \draw [->,line width=0.2mm,domain=1:-1] plot ({\x+ 1.3}, {-1*\x});
    \node (A) at (0.8,0.3) {(L)};
    \node (A) at (0.8,0.1) {$U>0$};
    \node (A) at (0.8,-0.05) {$V<0$};
    \node (B) at (1.8,0.3) {(R)};
     \node (A) at (1.8,0.1) {$U<0$};
    \node (A) at (1.8,-0.05) {$V>0$};
    \node (C) at (1.25,0.8) {(F)};
    \node (E) at (1.25,0.6) {$U >0$};
     \node (F) at (1.25,0.45) {$V >0$};
    \node (D) at (1.3,-0.4) {(P)};
    \node (D) at (1.3,-0.6) {$U <0$};
    \node (D) at (1.3,-0.75) {$V <0$};
    \node [font=\small] (U0) at (2.6,1.) {$U=0$};
    \node [font=\small] (V0) at (0,1. ) {$V=0$};
     \draw[ mblue, thick] (2.35,0.9) to[out=-120,in=120] (2.35,-0.9)
      node[midway, above, inner sep=2mm] {};
      \draw[ mblue, thick] (0.25,0.9) to[out=-60,in=60] (0.25,-0.9)
      node[midway, above, inner sep=2mm] {};
      \draw[ mred, thick, decorate,decoration=zigzag] (2.15,1.1) to[out=-160,in=-20] (0.45,1.1)
      node[midway, above, inner sep=2mm] {};
       \draw[ mred, thick, decorate,decoration=zigzag] (2.15,-1.1) to[out=160,in=20] (0.45,-1.1)
      node[midway, above, inner sep=2mm] {};
    \end{tikzpicture}
    \caption{Different regions using lightcone coordinates $U,V$ which reflect the Kruskal extension of the black hole geometry.}
    
    \label{fig:rindler}
\end{figure}
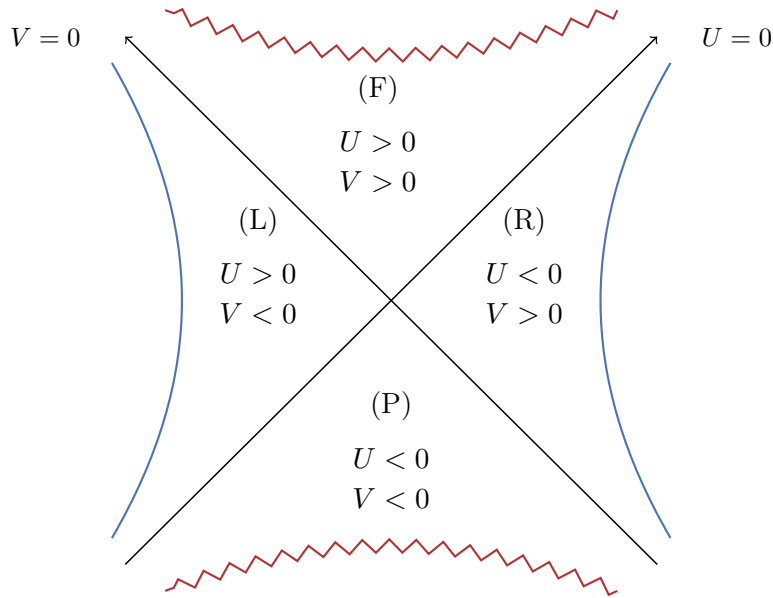

We work with the black hole metric \eqref{eq:pbhmetric} in the exterior region (Region R) and analytically extend it into other regions. To do this, we define the Tortoise coordinate $z_*$ given by
\beq
z_* = \int_0^z {{\rm d}z\over f(z)} = z \, _2F_1\left(1,\frac{1}{d};1+\frac{1}{d};{z^d \over z_\mathrm{h}^{d}}\right)
\eeq
The Tortoise coordinate has a branch cut at $z = z_h$, using which we perform analytic continuation to other regions to give us the Kruskal extension of the exterior geometry. We use this to define the Rindler lightcone coordinates, which cover the maximally extended solution including the interior (Region F). In the exterior region, the coordinates are given by
\beq \label{lightcone}
U = -\exp \frac{-2\pi (T-z_*)}{\beta}; \quad V = \exp \frac{2\pi (T+z_*)}{\beta}
\eeq
For the interesting cases of planar BTZ black hole and the AdS$_5$ black brane, the asymptotic boundaries are at $UV =-1$ and $UV = -e^\pi$ respectively. 

Our discussion is essentially the same as in the Rindler example Appendix \ref{2drindler}. Using the above coordinate system in the exterior, we can venture into the future wormhole Region F by a straightforward analytic continuation which is specified using the physics of a future-directed, radially infalling geodesic $V =$ constant. This necessitates the analytic continuation of the Tortoise coordinate about $z_h$ which also analytically continues the time $T$ as follows:
\beq \label{ancontgeom1}
z_*  \mapsto z_* + \ic {\beta \over 4}, \qquad T  \mapsto T - \ic {\beta \over 4}, 
\eeq
This makes both $U, V > 0$. For the cases of BTZ black hole and the AdS$_5$ black brane, the singularities are at $UV =1$. We refer the reader to Appendix \S \ref{2drindler} for a thorough discussion of the modes close to the horizon, which factorize to a two-dimensional Rindler description. We also note that in three dimensions, it is also useful to study this and the subsequent analytic continuations using isotropic coordinates, and the analytic continuations in the isotropic radial coordinate in \cite{Chakravarty:2024bna} to go behind the horizon is equivalent.

Relating the right exterior coordinates (Region R) to the other exterior (Region L) involves performing a reverse analytic continuation from the interior to the exterior for the Tortoise coordinate. We again consider a null radial geodesic from the interior to the left exterior. 

Combining everything together, from the perspective of the right exterior, going to the left exterior involves a net continuation in the time coordinate while we obtain a net zero imaginary part of the $z_*$ coordinates while going from the right exterior into the shared interior and coming out on the left exterior.
\beq \label{ancontgeom2}
T  \mapsto T - \ic {\beta \over 2},\qquad {\rm Im} \ z_* \sim 0
\eeq
The above operation takes us from the region $U<0, V>0$ to the region with $U>0, V<0$. The (real) left boundary time is related to the bulk time as $t_L = -T_L + c$, where we set $c =0$. The analytic continuation above essentially relates the real CFT times as $t_R  \mapsto -t_L - \ic {\beta \over 2}$. 

\subsection{The transformation and operator dictionary}

We are interested in extracting the right-moving modes near the bulk point using an analytic continuation of future operators which is equivalent to the action of left CFT operators. We use $O_L (O_R)$ to denote left (right) CFT operator insertions in position space.
We implement their extraction using Fourier transforms on the left boundary operators. 
\beq \label{leftopdef}
\begin{split}
O_{y,q,\sigma} =\int {\rm d}^d\delta y \, e^{-\ic q \cdot \delta y - {\delta y^2 \over 2 \sigma^2}}O_L({y+\delta y}), \qquad
O^\dagger_{y,q,\sigma} =\int {\rm d}^d\delta y \, e^{\ic q \cdot \delta y - {\delta y^2 \over 2 \sigma^2}}O_L({y+\delta y})
\end{split}
\eeq
Here  we have taken $\beta \, \omega \gg 1, \sigma > {\beta \over 2}  $, while the label $y$ denotes the hyperboloid on the left boundary. 
This suffices to extract the late-time phases near the bulk point which can be used to perform the scattering experiments.

\subsubsection{Analytic continuation}
Using the behaviour of the modes in the Kruskal geometry, we now motivate an analytic continuation from the one-sided past-future experiment to a two-sided left-right experiment which gives a bulk point in the interior.

To conveniently extract the oscillator modes, let us define the smearing factor given by:
\beq \label{eq: smearing function}
f_\beta(\delta t, \delta t') = {\beta \over 2 \pi} { e^{{\beta \omega \over 2}} \over (\delta t' + \delta t)^2 + {\beta^2 \over 4}}
\eeq
Note that here we have boosted up the strength by explicitly multiplying with factors of $e^{{\beta \omega \over 2}}$, in order to get rid of the suppressions later on and to make the measurement $\mathcal O(1)$. In the complex $\delta t'$ plane, the function $f_\beta(\delta t, \delta t')$ has poles at $\delta t' = -\delta t \pm \ic {\beta /2}$. We use the poles of the above function are important to take the operator insertions from the future hyperboloid to the left CFT.

Using the definition in \eqref{leftopdef} and the function \eqref{eq: smearing function}, we can extract the oscillator modes near the bulk point in terms of future right CFT operators as given below:
\beq
\begin{split} \label{finform}
O_{Ly,q,\sigma} &=\int {\rm d}^d\delta y \, e^{-\ic q \cdot \delta y - {\delta y^2 \over 2 \sigma^2}}O_L({y+\delta y})\\
&= \int {\rm d}^d\delta y  \ e^{-\ic {\bf q}.{\bf \delta y} - {({ \delta y})^2\over 2 \sigma^2} } \int {{\rm d}(\delta t')} \ f_\beta(\delta t, \delta t') \ e^{-\ic \omega \delta t'} \ O_R(t + \delta t',{\bf y+\delta y}) \\
O^\dagger_{Ly,q,\sigma} &=\int {\rm d}^d\delta y \, e^{\ic q \cdot \delta y - {\delta y^2 \over 2 \sigma^2}}O_L({y+\delta y})\\
&= \int {\rm d}^d\delta y  \ e^{+\ic {\bf q}.{\bf\delta y} - {(\delta { y})^2\over 2 \sigma^2} } \int {{\rm d}(\delta t')} \ f_\beta(\delta t, \delta t') \ e^{\ic \omega \delta t'} \ O_R(t +\delta t',{\bf y+\delta y})\\
\end{split}
\eeq

While performing the integral over $\delta t'$, we have closed the contour in the lower half plane when the mode $e^{-\ic \omega \delta t'} $ is involved and in the upper half plane for $e^{+\ic \omega \delta t'}$. We also used the fact that for late time absorption operators, the left and the right boundary times are related by the continuation given by $t_R  \mapsto -t_L -{\ic \beta \over 2}$ which takes the operator insertions from $O_R(y) \to O_L(y)$. This is similarly generalized to $t_R  \mapsto -t_L +{\ic \beta \over 2}$ for late time emission operators.

An important point is that the analytic continuation takes the positive (negative) modes from the right future hyperboloid to negative (positive) frequency modes on the left CFT hyperboloid. Under this action, since the bulk point is shifted to the interior, the shapes of the hyperboloids is not preserved (see \S \ref{sec:cauhyp}). In the bulk however, since $t_L = -T_L$, the transformation maps relates the Unruh mode, i.e. maps the positive (negative) component on the right to the positive (negative) component on the left.

As explained, we also obtain an overall factor of $e^{-\beta \omega \over 2}$ for both the modes (see also \cite{Basha:2018bvi}), which is taken into account by the exponential factor $e^{{\beta \omega \over 2}}$ in the smearing function \eqref{eq: smearing function}. For the case $\beta \, \omega \gg 1$, this implies we need to highly boost up the signal while shifting the bulk point from the exterior to the interior, else a naive analytic continuation gives us a suppressed signal for the bulk point irrespective of the oscillator mode\footnote{As we expect, extracting details of the interior should be costlier compared to the exterior!}. The expressions for the modes in \eqref{finform} provide us with the necessary mode-decomposition, which results from the analytic continuation of a well-defined experiment in the right exterior region. 
\subsubsection{Local bulk modes}
Similar to the smeared modes on the right CFT, the propagators involving left smearing operators $O_{Ly,q,\sigma} / O^\dagger_{Ly,q,\sigma}$ defined in \eqref{leftopdef} also take the form of a WKB phase.  
To venture to the interior using the left moving modes, we can extend the Hamilton-Jacobi function $S(p,X)$ using the continuation from the right exterior to the interior; and the function $S(q,X)$ using  the continuation from the left exterior to the interior region. The bulk point is picked by the intersection of geodesics in the black hole interior shot from the hyperboloids.

Similar to the exterior case, using a plane wave expansion for the propagator, we can read off the local oscillators in the following form:
\begin{equation}\label{eq:dictionary1}
\begin{split}
  O^\dagger_{p,\sigma} \ket{\Psi } \;&=\; 
        \int \ls {\rm d} P \, \psi^+ \rs \ a^\dagger_{X,P}  \ket{\Omega},   \mbox{if $O^\dagger$ has support on past hyperboloid}, \\
   O_{L,p,\sigma} \ket{\Psi }  \;  &= \; \int \ls {\rm d} P \, \psi^- \rs \ \tilde{a}^\dagger_{X,P} \ket{\Omega},  \mbox{if $O^\dagger$ has support on the future hyperboloid}.
\end{split}
\end{equation}
Here $O_{L,p,\sigma}$ are operators obtained by acting upon $O^\dagger_{p,\sigma}$ with the integral transform \eqref{finform}, which in turn introduces local right-moving modes near $X$.  This generalizes the exterior dictionary of \cite{Caron-Huot:2025hmk, Caron-Huot:2025she} to the interior, where we showed its plane wave limit in Table \ref{movers}. It is an important open question to generalize this dictionary to general wormhole geometries.

In particular, our operator dictionary has clear parallels with the state-dependent dictionary of \cite{Papadodimas:2012aq, Papadodimas:2015jra, Verlinde:2013qya, Verlinde:2013uja, Nomura:2012cx}, where the important lesson is that operators depend upon the state. This is \textit{mildly} reflected in our case, the operators depend for instance upon the integral transform which needs to pole at the correct inverse temperature $\beta$. 

\subsection{Interior experiments using thermal correlator}
We now consider experiments using the four point correlator on a single / two timefold contours. We start by outlining the domain of validity of our experiment and then discuss them in detail.

\subsubsection{Allowed timescale for interior experiments} 
\label{sec:twoptsad}
Let us firstly outline the domain of validity of our signal. We work with the correlators such that the time differences between initial and final operator insertions is considerably large enough.\footnote{As opposed to our $N=\infty$ limit, the expectation at finite $N$ is that the time differences between initial and final operator insertions cannot be very large since beyond the scrambling timescale, the backreaction becomes very large which breaks our bulk description. To obtain sensible answers, we perform measurements of the correlator  at intermediate times, i.e. 
\beq
\Delta t = \abs{t_{i} - t_{f}} <\mathcal{O} \lc t_{\rm scr} \rc.
\eeq
Using the boost isometry, we fix the early-time insertions at $t_i=0$. In the case of the BTZ geometry, we have a valid description of the future interior in the time range $t_f <\mathcal{O} \lc t_{\rm scr} \rc$. The transformation \eqref{finform} takes this to the left boundary where this corresponds to $\abs{t_L} <\mathcal{O} \lc t_{\rm scr} \rc$.} 

These effects become particularly important in $d >2$. In the AdS$_5$ black brane case, the singularities are pulled inwards, and the effects should be visible at the level of the two-point function \cite{Fidkowski:2003nf, Klosch:1995qv}. Upon the bulk point approaching the singularity, the four point function is instead is dominated by products of two-point functions which have lightcone singularities on the second sheet \cite{Fidkowski:2003nf, Festuccia:2005pi, Ceplak:2024bja, Afkhami-Jeddi:2025wra}.
 Therefore for genuine contributions from scattering using the connected four-point function (and not from bouncing geodesics which are disconnected part of the four point function), we need:
\beq
\Delta t = \abs{t_{i} - t_{f}} \gg {\beta \over 2}
\eeq
It is no surprise why this new timescale appears, since the Fourier transform of the Wightman two-point function receives contribution from the bouncing geodesic saddles from branch cuts at complex time $t = \lc \pm \frac{\beta}{2} + \frac{\ic \beta}{2}\rc$ and its multiples (see Fig 3 of \cite{Afkhami-Jeddi:2025wra}). Note here that this works since the extrapolate limit of the bulk-to-boundary Wightman propagator that we use gives us the same boundary Wightman function as in \cite{Afkhami-Jeddi:2025wra}. For the four-point correlator to give a definitive signal, we must stay away from this region with the real part of time much larger than ${\beta \over 2}$.

Finally we discuss how the connected part behaves as the bulk point approaches close to the singularity in \S \ref{sec:bhsing}. There we see that the signal becomes faint as the bulk point approaches the singularity, i.e. its intensity goes to zero. Consequently in this regime  the disconnected part of the four-point function, i.e. contribution from the two point function becomes dominant which is a better probe of the singularity.

\paragraph{Competing saddles:} Another competing saddle can appear at intermediate times when one of the wavepackets' energy is much larger than the other, for instance when $\omega_4 \gg \omega_2$. We arrive at this new saddle provided that the suppression factor coming from the Fourier transform dominates over our four-point saddle, i.e.
\beq
\exp\lc -{ \beta \omega_2 \over 2}-{{1 \over \sigma_2^2}\lc t_4 + \ic {\beta / 2} - t_2\rc^2}\rc \gg \exp \lc -{\beta \over 2} (\omega_2 + \omega_4)\rc 
\eeq
where $t_4$ and $t_2$ are the (real) boundary times for the corresponding operator insertions. In this limit, the contribution is the factorized contribution of two two-point correlators.

\paragraph{Imaging the past interior:} Let us relax to the case where the operator insertions at $t_f$ can also appear earlier than $t_i$. Consider the initial time of the operator insertions as $t_i \sim 0$. Then there is no sharp signal in the following temporal range, since here we need to take into account additional contributions to the four-point correlator coming from two-point functions:
\beq
-\frac{\beta}{2} < t_f < \frac{\beta}{2}.
\eeq
For $t_f<-\frac{\beta}{2}$, since we work with the Wightman functions, our smearing prescription automatically takes us to scattering events in the vicinity of a bulk point in the past black hole interior.

\begin{figure}
    \begin{subfigure}{0.5\textwidth}
    \centering
\begin{tikzpicture}[x=0.75pt,y=0.75pt,yscale=-1.15,xscale=1.15]
\draw    (176,185.5) -- (252,263) ;
\draw    (251,109) -- (175.5,186.5) ;
\draw [thick]   (254,245) .. controls (257,223) and (305,210) .. (307,248) ;
\draw  [thick]  (254,143) .. controls (267,167) and (293,175) .. (305,143) ;
\draw [thick, color=mblue  ,decoration={markings,mark=at position 0.5 with \arrow{<}},postaction=decorate ,draw opacity=5 ]   (242,197) -- (264,231) ;
\draw [thick, color=mblue  ,decoration={markings,mark=at position 0.65 with \arrow{<}}, postaction=decorate  ,draw opacity=5 ]   (242,197)  -- (227,217) ;
\draw [thick, color=mblue!50  ,decoration={markings,mark=at position 0.35 with \arrow{<}}, postaction=decorate  ,  ,draw opacity=5 ]   (297,227) .. controls (287.16,220.95) and (277,214) .. (270,210) .. controls (263,206) and (213.13,177.89) .. (218,186) ;
\draw [thick, color=mblue  ,decoration={markings,mark=at position 0.35 with \arrow{>}}, postaction=decorate  ,draw opacity=5 ]   (242,197) -- (261.67,179.45) -- (279,164) ;
\draw [thick,  color=mblue!50  ,decoration={markings,mark=at position 0.65 with \arrow{<}}, postaction=decorate  ,draw opacity=5 ]   (227,216) .. controls (224,215) and (224,215) .. (265,157) ;
\draw [thick, color=mblue  ,decoration={markings,mark=at position 0.5 with \arrow{<}}, postaction=decorate  ,draw opacity=5 ]   (218,186) -- (242,197) ;
\draw[dashed]    (231,171) -- (207,197) (231,221) -- (207,201) ;
\draw (252,190) node [anchor=north west][inner sep=0.75pt]  [font=\small] [align=right] {$X$};
\end{tikzpicture}
\caption{}
\end{subfigure}
\begin{subfigure}{0.5\textwidth}
    \centering
\begin{tikzpicture}[scale = 1.5]
\draw[dashed] (-0.7,0.3)--(-0.3,0.7) (-0.65,-0.35) -- (-0.45,-0.55);
\draw[dashed, thick, orange](0.7,0.3) --(0.3,0.7);
\draw [thick, color=mblue  ,decoration={markings,mark=at position 0.5 with \arrow{>}}, postaction=decorate  ,draw opacity=5 ]   (1,-1)--(0,0);
\draw [thick, color=mblue  ,decoration={markings,mark=at position 0.5 with \arrow{>}}, postaction=decorate  ,draw opacity=5 ] (0,0) ..controls (-0.65,0.35) ..(-0.6,0.4) ;
\draw [thick, color=mblue!50  ,decoration={markings,mark=at position 0.5 with \arrow{<}}, postaction=decorate  ,draw opacity=5 ]   (1.4,-0.7)--(-0.6,0.4) ;
\draw (0,-0.05) node [below] {$X$};
\draw [thick, color=mblue  ,decoration={markings,mark=at position 0.5 with \arrow{<}}, postaction=decorate  ,draw opacity=5 ] (0,0) ..controls (-0.55,-0.45) ..(-0.6,-0.4) ;
\draw [thick, color=mblue!50  ,decoration={markings,mark=at position 0.35 with \arrow{<}}, postaction=decorate  ,draw opacity=5 ]   (-0.6,-0.4)--(0.4,0.6) ;
\draw [thick, color=mblue  ,decoration={markings,mark=at position 0.5 with \arrow{>}}, postaction=decorate  ,draw opacity=5 ] (-1.2,-0.7) ..controls (0.35,0.65) ..(0.4,0.6) ;
\draw [thick, color=mblue  ,decoration={markings,mark=at position 0.5 with \arrow{>}}, postaction=decorate  ,draw opacity=5 ] (0,0) ..controls (0.5,0.5) ..(0.55,0.45) ;
\draw [thick, color=mblue!50  ,decoration={markings,mark=at position 0.7 with \arrow{>}}, postaction=decorate  ,draw opacity=5 ]   (0.55,0.45)-- (-0.7,-1) ;
\filldraw (0,0) circle (0.75pt);
\draw (-2,0.5) -- (0,-1.5) -- (2,0.5);
\end{tikzpicture}
\caption{}
\end{subfigure}
    \caption{(a) Exterior experiment with two timefolds with the contour \eqref{bp2fold}. (b) Local behaviour near an interior bulk point upon applying transforms which introduces additional foldings on the top right (dashed orange line).}
    \label{fig:2to2int}
\end{figure}
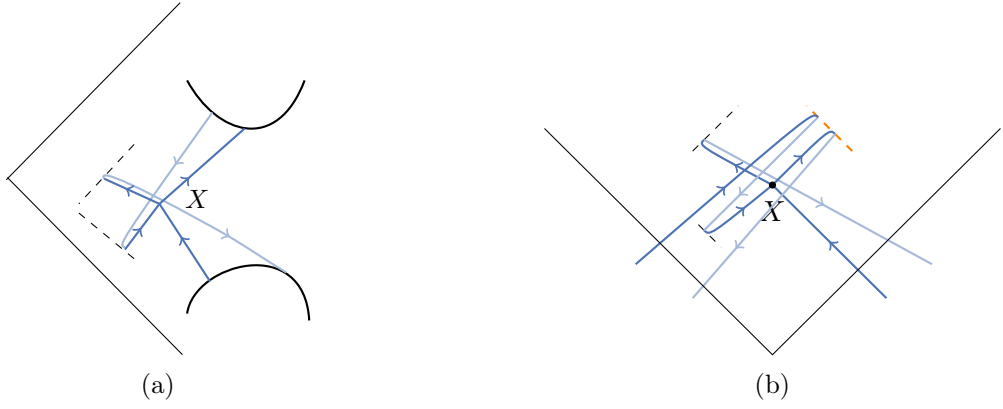

\subsubsection{Radar experiment for black hole interior}

Since we have discussed the experiments in some detail in the exterior and also their boundary description for planar BTZ interior in the earlier sections, we will simply translate to the general discussion without repeating many details. We consider the four-point correlator $\langle \Psi |  \, O_{x_4, p_4, \sigma_4} \, O_{Ly_3, q_3, \sigma_3}\, O^\dagger_{x_2, p_2, \sigma_2} \, O^\dagger_{x_1, p_1, \sigma_1} \,  | \Psi \rangle $ where we have used the transform \eqref{finform} to take $y_3$ to the left side. Thus this gives us an exterior measurement which is mathematically equivalent to a bulk process taking place in the interior.

Let us now try to understand bulk locality in this context. We specialize to the setting where we shoot the wavepackets towards the bulk point $X$. For us to obtain the flat space limit near a bulk point, we need the wavepackets to not miss each other and to satisfy local bulk momentum conservation $\sum_i P_i = 0$. A failure to satisfy either constraint leads instead to a general Gaussian in Appendix \ref{sec:wavepackets} (see \cite{Caron-Huot:2025hmk}).

It is only when both conditions are met that the Gaussian approaches a delta function. The four-point amplitude then factorizes to a $3 \to 1$ bulk amplitude:
\beq
\langle \Psi |  \, O_{x_4, p_4, \sigma_4} \, O_{Ly_3, q_3, \sigma_3}\, O^\dagger_{x_2, p_2, \sigma_2} \, O^\dagger_{x_1, p_1, \sigma_1} \,  | \Psi \rangle \quad  \mapsto \quad  \langle \ a_{X,P_4} \, \tilde{a}_{X,Q_3} \,  a^\dagger_{X,P_2} \, a^\dagger_{X,P_1}  \ \rangle
\eeq
about an interior bulk point $X$. Note that we follow contour ordering for the transformed boundary correlator as for the original correlator.

 The factorization to the $3 \to 1$ bulk amplitude is visually depicted below, where the bulk operator ordering remains the same as for the boundary correlator.
\beq
    \centering
    \begin{tikzpicture}[scale=0.8] 
    \draw  (8,0.1)--(8,0) ..controls (12,0.1) and (12.5,1.1).. (13.5,1.0) -- (13.5,-0.3) -- (8,-0.3)--(8,-0.4);
\node [below] at (9.5,-0.3) {$O_{x_4, p_4, \sigma_4}$};
\node [above] at (12.5,0.95) {$O^\dagger_{Ly_3, q_3, \sigma_3}$};
\node [above] at (8.6,0.2) {$O^\dagger_{x_1, p_1, \sigma_1}$};
\node [above] at (10.4,0.35) {$O^\dagger_{x_2, p_2, \sigma_2}$};
\filldraw (13,0.95) circle (2pt); \filldraw (9.5,-0.3) circle (2pt); \filldraw (9,0.04) circle (2pt); \filldraw (10,0.15) circle (2pt);
\draw (15,0.35) -- (15,-0.15) -- (15.5,-0.15);
\node [above] at (15.25,-0.15) {$t$}; 
\node [right] at (15.25,0.15) {$\quad  \mapsto \, \, $}; 
    \end{tikzpicture}
    \qquad \quad
     \begin{tikzpicture}[scale=0.8] 
\draw  (0,0.1)--(0,0) -- (5.5,0) -- (5.5,-0.3) -- (0,-0.3)--(0,-0.4);
\node [below] at (1.2,-0.3) {$a_{X, P_4}$};
\node [above] at (4.5,-0) {$\tilde{a}_{X, Q_3}$};
\node [above] at (0.6,0) {$a^\dagger_{X, P_1}$};
\node [above] at (2.4,-0) {$a^\dagger_{X, P_2}$};
\filldraw (4.5,0) circle (2pt); \filldraw (1.5,-0.3) circle (2pt); \filldraw (1,0) circle (2pt); \filldraw (2,0) circle (2pt);
\draw (6.5,0.35) -- (6.5,-0.15) -- (7,-0.15);
\node [above] at (6.85,-0.15) {$z_*$}; 
\end{tikzpicture}
\eeq
The future timefold introduces a sum over the out states, which is represented by the dashed line in Fig \ref{mainfig00}(a). One crucial difference here from the exterior case is that the local modes are now defined in terms of the local timelike Killing vector $z_*$ instead of the asymptotic Killing vector $T$. The WKB phase takes care of the bulk evolution from the hyperboloids to the bulk point.

Using the operator dictionary after the transform, we obtain a factorization formula as in \eqref{rep2} for local scattering amplitudes about a bulk point located in the black hole interior.
\beq 
\begin{split}
\langle \Psi |  &\, O_{x_4, p_4, \sigma_4} \, O^\dagger_{Ly_3, q_3, \sigma_3}\,  O^\dagger_{x_2, p_2, \sigma_2} \, O^\dagger_{x_1, p_1, \sigma_1} \,  | \Psi \rangle \\
 & \approx \int \qty[\prod \diff P_i\,\psi_i] \qty[ \diff Q_3\,\psi_3]  \sqrt{-g} \, (2\pi)^{d+1}\ \delta^{d+1}\qty(\textstyle{P_1 + P_2 - P_4 - Q_3})\, \ \ic \cM(\{P_i, Q_3\}), \\
\end{split}
\eeq
Here $\cM(\{P_i, Q_3\})$ denotes the $3\to 1$ scattering amplitude on our in-in contour.

It would be an interesting exercise to understand the relation of the present discussion to the hydrodynamic regime where non-trivial time ordering prescriptions for Schwinger-Keldysh have been proposed (see \cite{Glorioso:2018mmw, deBoer:2018qqm, Chakrabarty:2019aeu, Bu:2020jfo , Loganayagam:2022teq, He:2022jnc, Loganayagam:2023pfb, Sivakumar:2024iqs, Martin:2024mdm, Loganayagam:2025ell} for related works exploring different contexts). The partition functions corresponding to these prescriptions wrap around the branch cut in Tortoise coordinate by an infinitesimal amount before going out to the exterior. Gradually reducing the energy of our setup leads to larger sized wavepackets and in turn we lose bulk locality.

\subsubsection{Four-point correlator on two timefolds}

 \begin{figure}[t!] 
    \centering
    \begin{tikzpicture}[scale=0.85]
    \draw  (0,0.1)--(0,0) -- (5.5,0) -- (5.5,-0.3) -- (0,-0.3)--(0,-0.6) -- (5.5,-0.6) -- (5.5,-0.9) -- (0,-0.9) --(0, -1);
\node [below] at (1.8,-0.8) {$O_{x_4, p_4, \sigma_4}$};
\node [above] at (4.5,-0) {$O_{y_3, q_3, \sigma_3}$};
\node [above] at (0.6,0) {$O^\dagger_{x_1, p_1, \sigma_1}$};
\node [below] at (4.5,-0.8) {$O^\dagger_{y_2, q_2, \sigma_2}$};
\filldraw (4.3,0) circle (2pt); \filldraw (4.6,-0.9) circle (2pt); \filldraw (1,0) circle (2pt); \filldraw (1.3,-0.3) circle (2pt);
\node [right] at (6.25,-0.55)  { $  \mapsto $};
\draw  (8,0.1)--(8,0) ..controls (10,0) and (11.5, 1.4).. (13.5,1.3) -- (13.5,-0.3) -- (8,-0.3)--(8,-0.6) -- (13.5,-0.6) -- (13.5,-1.9)..controls (11.5,-2) and (10,-0.8).. (8,-0.9)--(8,-1);
\node [below] at (9.2,-1) {$O_{x_4, p_4, \sigma_4}$};
\node [above] at (12.5,1.55) {$O^\dagger_{Ly_3, q_3, \sigma_3}$};
\node [above] at (8.6,0.2) {$O^\dagger_{x_1, p_1, \sigma_1}$};
\node [below] at (12.7,-2.1) {$O_{Ly_2, q_2, \sigma_2}$};
\filldraw (12.9,-1.9) circle (2pt); \filldraw (8.8,-0.3) circle (2pt); \filldraw (12.7,1.28) circle (2pt); \filldraw (8.5,0.0) circle (2pt);
\draw[dashed, thick] (13.5,1.3)--(10.5,1.3) (13.5,-1.9) -- (10.5, -1.9) ;
\draw[<->] (14.5,1.3) -- (14.5, -1.9);
\node [right] at (14.6,-0.2) {$\beta$};
    \end{tikzpicture}
    \caption{The transformation in boundary time to a deformed boundary contour which takes us to Fig \ref{fig:2to2int}, where the dashed lines indicate the thermal circle identification $\tau \sim \tau + \beta$.} \label{fig:otocexp}
\end{figure}
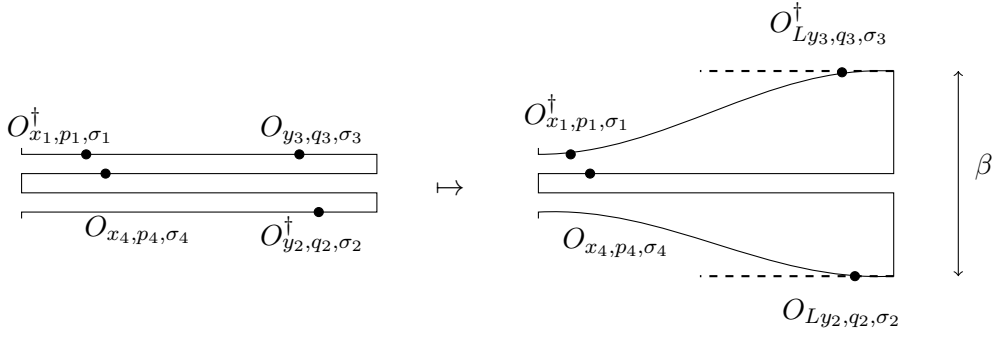

Most of the things here are similar to the radar discussion, so we will skip details here stated elsewhere in the text. 

\paragraph{Exterior:} At the level of the four-point correlator, we can consider the experiment shown in Fig \ref{redpaths} in the black hole exterior. Crossing the lightcones appropriately take us to the bulk point discontinuity in the exterior. 

This has one emission and one absorption operator each in the past and in the future. For a configuration to give us a scattering amplitude, we need to put the operators on a contour using two timefolds.
\beq \label{mtfiiex1}
\begin{split}
\langle \Psi | & \, O^\dagger_{y_2, q_2, \sigma_2} \, O_{x_4, p_4, \sigma_4} \, O_{y_3, q_3, \sigma_3}\,   O^\dagger_{x_1, p_1, \sigma_1} \,  | \Psi \rangle \\
 \end{split}
\eeq
Each timefold introduces a sum over out states.  In the bulk, this corresponds to the following scattering amplitude:
\beq \label{mtfiiex2}
\begin{split}
\langle  \, b^\dagger_{X,Q_2} \, a_{X,P_4} \, b_{X,Q_3}\,   a^\dagger_{X,P_1} \,   \rangle  = \sum_{\rm out} \langle \Omega   |\, b^\dagger_{X,Q_2} | {\rm out}_1\rangle \, \langle {\rm out}_1 | a_{X,P_4} \,|  {\rm out}_2 \rangle \, \langle {\rm out}_2  |\, b_{X,Q_3} \,   a^\dagger_{X,P_1} \, | \Omega \rangle
\end{split}
\eeq
 where we have now introduced the completeness relation over the space of out states twice, which corresponds to the two future timefolds.

\paragraph{Interior:} Let us consider a left-right correlator given below which lies on two timefolds.
\beq \label{4pt2foldex}
\langle \Psi |  \, O_{Ly_2, q_2, \sigma_2} \, O_{x_4, p_4, \sigma_4}\,  O^\dagger_{Ly_3, q_3, \sigma_3} \, O^\dagger_{x_1, p_1, \sigma_1} \,  | \Psi \rangle
\eeq
The left-right correlator has a similar decomposition in terms of left-right amplitudes as in \eqref{mtfiiex1}. The left-right can be tuned to locate a bulk point in the interior, where we obtain the following $2 \to 2$ bulk amplitude $\langle  \, \tilde{a}^\dagger_{X,Q_2} \, a_{X,P_4} \, \tilde{a}_{X,Q_3}\,   a^\dagger_{X,P_1} \,   \rangle$.
\beq \label{bp2fold}
    \centering
    \begin{tikzpicture}[scale=0.8] 
\draw  (8,0.1)--(8,0) ..controls (10,0) and (11.5, 1.4).. (13.5,1.3) -- (13.5,-0.3) -- (8,-0.3)--(8,-0.6) -- (13.5,-0.6) -- (13.5,-1.9)..controls (11.5,-2) and (10,-0.8).. (8,-0.9)--(8,-1);
\node [below] at (9.2,-1) {$O_{x_4, p_4, \sigma_4}$};
\node [above] at (12.5,1.55) {$O^\dagger_{Ly_3, q_3, \sigma_3}$};
\node [above] at (8.6,0.2) {$O^\dagger_{x_1, p_1, \sigma_1}$};
\node [below] at (12.7,-2.1) {$O_{Ly_2, q_2, \sigma_2}$};
\filldraw (12.9,-1.9) circle (2pt); \filldraw (8.8,-0.3) circle (2pt); \filldraw (12.7,1.28) circle (2pt); \filldraw (8.5,0.0) circle (2pt);
\draw[dashed, thick] (13.5,1.3)--(10.5,1.3) (13.5,-1.9) -- (10.5, -1.9) ;
\draw[<->] (14.5,1.3) -- (14.5, -1.9);
\node [left] at (14.4,-0.2) {$\beta$};
\draw (15,0.35) -- (15,-0.15) -- (15.5,-0.15);
\node [above] at (15.25,-0.15) {$t$}; 
\node [right] at (15.25,-1.15) {$\quad  \mapsto \, \, $}; 
    \end{tikzpicture}
    \qquad \quad
     \begin{tikzpicture}[scale=0.8] 
\draw  (0,0.1)--(0,0) -- (5.5,0) -- (5.5,-0.3) -- (0,-0.3)--(0,-0.4);
\draw  (0,0.1)--(0,0) -- (5.5,0) -- (5.5,-0.3) -- (0,-0.3)--(0,-0.6) -- (5.5,-0.6) -- (5.5,-0.9) -- (0,-0.9) --(0, -1);
\node [below] at (1.8,-0.8) {$a_{X, P_4}$};
\node [above] at (4.5,-0) {$\tilde{a}_{X, Q_3}$};
\node [above] at (0.6,0) {$a^\dagger_{X, P_1}$};
\node [below] at (4.5,-0.8) {$\tilde{a}^\dagger_{X, Q_2}$}; 
\filldraw (4.3,0) circle (2pt); \filldraw (4.6,-0.9) circle (2pt); \filldraw (1,0) circle (2pt); \filldraw (1.3,-0.3) circle (2pt);
\end{tikzpicture}
\eeq
The transformed correlator and the local scattering amplitude in the  interior are shown above where we follow contour ordering for the transformed boundary correlator. As in the radar case, one can again obtain a similar factorization formula
\beq 
\begin{split}
\langle \Psi |  &\,  O_{Ly_2, q_2, \sigma_2} \, O_{x_4, p_4, \sigma_4} \,O^\dagger_{Ly_3, q_3, \sigma_3}\,   O^\dagger_{x_1, p_1, \sigma_1} \,  | \Psi \rangle \\
 & \approx \int \qty[\prod \diff P_i\,\psi_i] \qty[ \diff Q_i\,\psi_i]  \sqrt{-g} \, (2\pi)^{d+1}\ \delta^{d+1}\qty(\textstyle{P_1-P_4 -Q_3 +Q_2})\, \ \ic \cM(\{P_i, Q_i\}), \\
\end{split}
\eeq
which now describes a $2 \to 2$ amplitude on an otoc with two timefolds that is specified using the delta function $\delta^{d+1}(P_1-P_4 -Q_3 +Q_2)$.

On a general note, revealing the interior requires sensible measurements, and the process hence is usually delicate. Some examples of this include state-dependent bulk to boundary maps proposed and associated physical observables for large AdS black holes which hold at the level of correlators \cite{Papadodimas:2012aq, Papadodimas:2013jku, Papadodimas:2015jra, Verlinde:2013qya, Verlinde:2013uja, Nomura:2012cx, deBoer:2018ibj, Chakravarty:2020wdm}. It should be noted that correlations involving smeared bulk modes are quite subtle, for instance very close to the horizon, and require a delicate treatment in any asymptotic measurement \cite{Raju:2018zpn, Raju:2020smc, Chakraborty:2021rvy}. While we cannot exhaustively list the literature on this, see also clean tools to reveal the interior in lower-dimensional models \cite{Kourkoulou:2017zaj, Brustein:2018fkr, Dhar:2018pii}, see \cite{Heemskerk:2012mn, Almheiri:2017fbd, Roy:2018ehv} for interior measurements using reconstruction techniques. 

Finally there also exist another class of precise asymptotic measurements which studies how information is localized in a theory of gravity, see \cite{ Laddha:2020kvp, Raju:2020smc, Chowdhury:2021nxw, Chowdhury:2020hse, Chowdhury:2021nxw, Chakraborty:2023los, Chakravarty:2023cll} for different instances where this holography of information manifests. It would be interesting to develop this further to connect with the perturbative experiments we perform in our work.

\section{Boundary hyperboloids for black hole interior}
\label{sec:cauhyp}

Boundary hyperboloids are boundary locus lightlike to a given bulk point. These are crucial in understanding aspects of bulk causality \cite{Caron-Huot:2025she}.  We are interested in understanding hyperboloids as a function of the bulk coordinates within the context of the black hole interior. 

Classically, the geodesics are specified by the Hamilton-Jacobi function $S(p, X)$. We solve for a geodesics with a boundary endpoint. The temporal coordinates are related by:
\beq \label{eqt}
    t - T = \pm \omega \int^{z}_0 \frac{{\rm d}z}{f(z) \sqrt{\omega^2  - {\bf p}^2 f(z)}} 
    \eeq
This follows from integrating the geodesic equation. Similarly, we obtain the following integral for the transverse positions:
\beq \label{eqy}
\mathbf{y} - \mathbf{X} =  {\bf p} \int^{z}_0 \frac{{\rm d}z}{\sqrt{\omega^2  - {\bf p}^2 f(z)}}
\eeq

\paragraph{Empty AdS example:} Here the boundary hyperboloids lightlike to the bulk point $X=(T,X^a,z)$ are given by:
\beq
    t -T = \frac{\pm \omega z}{\sqrt{\omega^2 - {\bf p}^2}}, \qquad y^a -X^a= \frac{p^a z}{\sqrt{\omega^2 - {\bf p}^2}}
\eeq
The above form is a parametric representation of the boundary coordinates lightlike to the bulk point, which also admits a more compact form.
\beq
    - (t - T)^2 + \delta_{ab} (y^a - X^a)(y^b - X^b) + z^2 = 0
\eeq
This is the equation for the boundary hyperboloid corresponding to a bulk point $X$ in empty AdS. We see that geometric features of the hyperboloid is characterized by the radial depth of the bulk point. 

\subsection{Hyperboloids from an interior point}
\label{hypeintpt}

\begin{figure}[t!]
    \centering
    \includegraphics[width=0.8\linewidth]{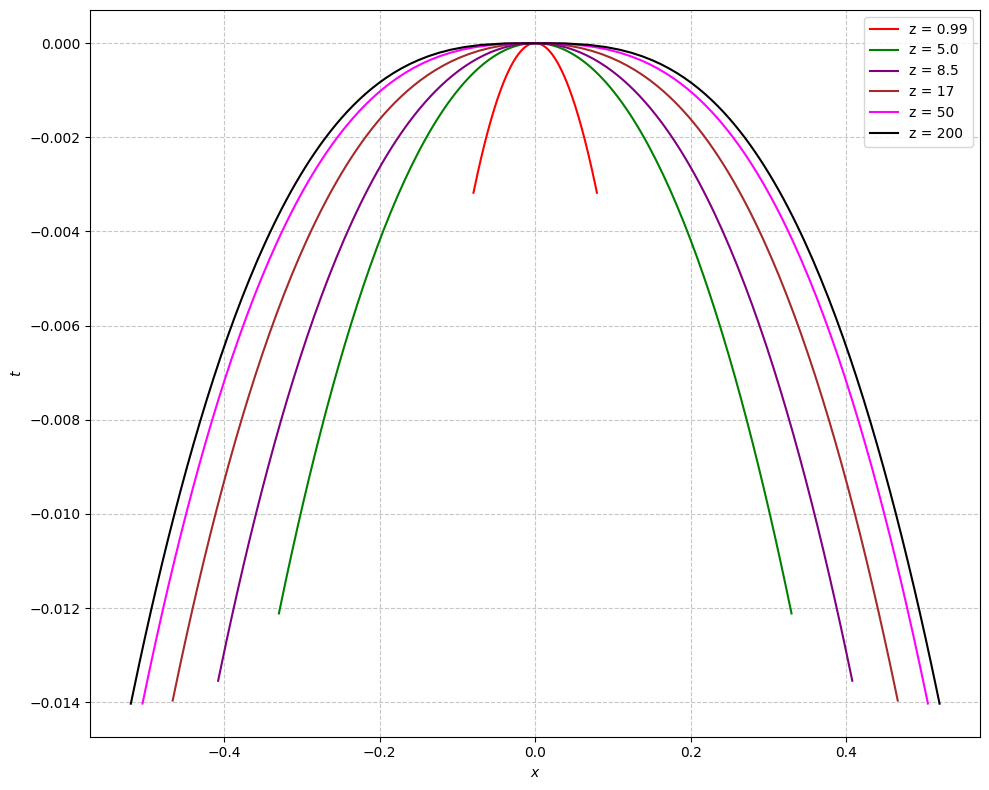}
    \caption{Boundary hyperboloids for the AdS$_5$ black hole corresponding to a radially infalling geodesic $T = z_*$, with $X=0$, plotted over a fixed range of shooting angle. The red line is just before the infalling geodesic crosses the horizon. }
    \label{fig:ads5sing}
\end{figure}

\paragraph{Planar BTZ:}In the BTZ case, there is only one transverse direction and momentum ${\rm p}$. For the planar BTZ black hole exterior region, we obtain the hyperboloid in equation \eqref{eq:planarbtzext}.
These equations capture the boundary hyperboloid when we are shooting from a bulk point in the black hole exterior $z < z_\mathrm{h}$. 

Let us now consider the case when we are in the black hole interior $z> z_\mathrm{h}$. Upon appropriate analytic continuation to push the bulk point into the interior, the hyperboloids now have the form in equation \eqref{eq:planarbtzint}.

\begin{figure}[t!]
    \centering
    \includegraphics[width=0.8\linewidth]{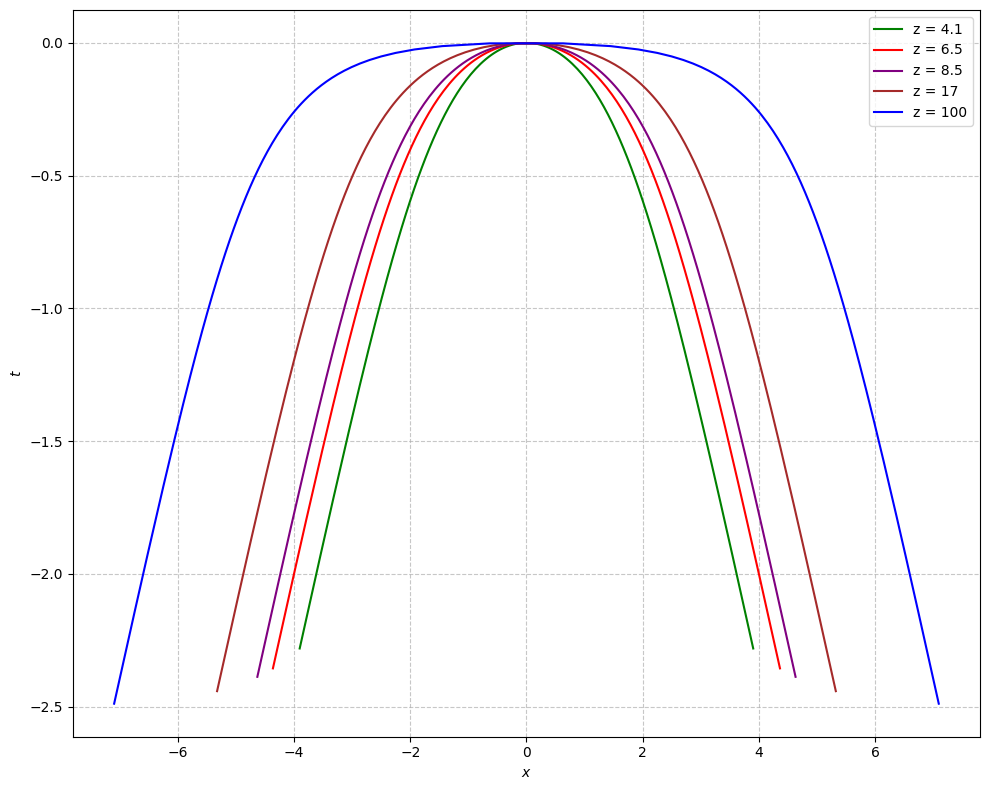}
    \caption{The hyperboloids for the BTZ black hole corresponding to a radially infalling geodesic $T = z_*$, with ${\rm X}=0$, over a shooting angular interval. The boundary hyperboloid asymptotically satisfies the equation $\abs{x} = \abs{t}$, with horizon radius $z_{\rm h} =1$.}
    \label{fig:btzhyp1}
\end{figure}

The above change has an interesting implication for the boundary hyperboloid: as shown in Fig \ref{fig:btzhyp1}, the hyperboloids start becoming flatter as we go further away from the horizon $z = z_\mathrm{h}$ and start approaching the singularity at $z \to \infty$. This is a feature arising upon approaching the black hole singularity. 

Again as before in empty AdS, we can utilize the parametric equations for the boundary hyperboloid to put the hyperboloids in a more compact equation. For the exterior regions, the equation for the boundary hyperboloids takes the form:
\beq
    -\omega^2 \coth^2 \frac{t - T}{z_\mathrm{h}} + {\rm p}^2 \coth^2 \frac{y - {\rm X}}{z_\mathrm{h}} = 0,
\eeq
while the corresponding equation with an interior bulk point takes the form:
\beq
    -\omega^2 \tanh^2 \frac{t - T}{z_\mathrm{h}} + {\rm p}^2 \coth^2 \frac{y - {\rm X}}{z_\mathrm{h}} = 0.
\eeq
While the above equations seem independent of the radial depth, obtaining a consistent solution of the above equation requires us to fix the radial depth, as reflected in the parametric form.

\paragraph{AdS$_{d+1}$ planar black hole: }
Finally let us look at the boundary hyperboloids for AdS$_{d+1}$ planar black holes. The solution to the temporal equation \eqref{eqt} gives us the equation:
\beq \label{tads5}
t-T = \pm \frac{\omega z \lc \sqrt{ \lc z \over z_{\rm h}\rc^d \frac{{\bf p}^2}{\omega^2-{\bf p}^2}+1} \rc \, F_1\left(\frac{1}{d};\frac{1}{2},1;1+\frac{1}{d}; \lc z \over z_{\rm h}\rc^d \frac{{\bf p}^2}{{\bf p}^2-\omega^2},\lc z \over z_{\rm h}\rc^d\right)}{\sqrt{{\bf p}^2 \left(\lc z \over z_{\rm h}\rc^d-1\right)+\omega^2}}
\eeq
Here the hypergeometric function is the Appell hypergeometric function. Similarly the solution to the transverse coordinate takes the form:
\beq
\mathbf{y} - \mathbf{X} = \frac{{\bf p} z \lc \sqrt{{\bf p}^2 \left(\lc z \over z_{\rm h}\rc^d-1\right)+\omega^2}\rc \, _2F_1\left(1,\frac{1}{2}+\frac{1}{d};1+\frac{1}{d}; \lc z \over z_{\rm h}\rc^d \frac{{\bf p}^2}{{\bf p}^2-\omega^2}\right)}{\omega^2-{\bf p}^2}
\eeq
Extending these formula from the exterior to the interior using the analytic continuation, we can check that that the AdS$_5$ boundary hyperboloids obey similar features as previously observed for the BTZ interior (see Fig \ref{fig:ads5sing}).

\subsection{The black hole singularity}
\label{sec:bhsing}
\paragraph{Exterior curvature:} The boundary hyperboloids reveal some interesting features as we approach the black hole singularity in AdS$_5$ Schwarzchild black holes. Using this, we write the exterior curvature of the boundary hyperboloid:
\begin{equation}
    \tilde{K}^{\mu \nu} = - {\rm p}_\rho \frac{\partial^2 y^\rho}{\partial {\rm p}_{\mu} \partial {\rm p}_{\nu}}= - \frac{\partial}{\partial {\rm p}_{\mu}} \qty({\rm p}_\rho \frac{\partial y^\rho}{\partial {\rm p}_{\nu}}) + \frac{\partial y^{\mu}}{\partial {\rm p}_{\nu}} = \frac{\partial y^{\mu}}{\partial {\rm p}_{\nu}}
\end{equation}
In the above expression, we separated the exterior curvature into a derivative acting on the normal condition, and an additional piece.
From \cite{Caron-Huot:2025she}, with a fixed bulk point $X$, the normal space to the hyperboloid $y^\mu(p,X)$ is specified by the momenta $n_\rho = {\rm p}_\rho$. The normal condition eliminates the first term within the brackets (since ${\rm p}_\rho \frac{\partial y^\rho}{\partial {\rm p}^{\nu}}=0$). 

A signature of the black hole singularity is that the exterior curvature of the boundary hyperboloids goes to zero at a small transverse angle. Since the momenta live in the normal space, the transverse momenta components $p_i$ go to zero as the bulk point approaches the singularity. However this could also arise as a feature of focal points of the Raychaudhuri equation. 

\paragraph{Signal strength:} We approach this problem using the coincidence limit of the correlator \eqref{4pt2foldex} with two past operator insertions and two future operator insertions.  As a diagnostic, a simple calculation tells us that in the high energy limit the magnitude of the correlator, i.e. its intensity very close to the singularity goes as follows:
\beq  \label{schint1}
\lim_{z \to \infty} I(z(\Delta t)) \propto  \begin{cases}
&{\rm constant}, \qquad  \, d =2 \\
&z^{-2} \log z, \qquad \, d =3\\
&z^{-2d+4},  \quad \qquad d \neq 3 \\
\end{cases}
\eeq
where $z$ is the radial location of the bulk point. Thus in $d>2$, we see that the signature of the black hole singularity is twofold: firstly the parallax going to zero, and secondly the intensity going to zero. This is expected from the signal, as the center of mass energy $s=2 g_{\mu \nu}P^{\mu} Q^{\nu}$ goes to zero when we approach the singularity. 

In particular, for the AdS$_5$ black brane, while there is an exponential dependence of $s$ with the boundary time away from the singularity, there is a sudden dip as the intersection point as we approach the singularity at $z \to \infty$. For $d=2$, the BTZ black hole singularity is not a true curvature singularity which leads to a constant value. Equation \eqref{schint1} shows that the intensity becomes constant near the singularity.

\subsubsection{Better signals from bouncing geodesics}
Since the signal from the connected part of the four-point correlator becomes faint, it is not useful anymore.
As pointed out in \S \ref{sec:twoptsad}, at timescales close to approaching the singularity we also receive contributions from saddles for the two-point function. Basically instead of the scattering picture described by the connected part of the four point correlator, the dominant contributions will be given by product of two-point functions which will pick up the bouncing geodesics. This follows from the GKPW decomposition of the four point correlator where each of the bulk to boundary propagator is Fourier transformed. This can also be understood from the extrapolate limit of correlators: the bulk-to-boundary Wightman propagator goes to the boundary two-point function, the Fourier transform helps pick up the bouncing geodesic saddle in real time since it is favorable to deform the contour to localize about the bouncing geodesic saddle \cite{Afkhami-Jeddi:2025wra}. 

Note also that the boundary hyperboloid description breaks down as the bulk point further approaches and comes very close to the singularity. This is because the high energy WKB approximation breaks down. As we can see from \eqref{schint1}, there is a dip in the overall correlator strength near the singularity along with the exterior curvature going to zero. In general, we would prefer a sharp jump rather than a smooth dip to diagnose the singularity, the absence of a signal is not a smoking gun! It may be so that we are still encountering a focal point of the Raychaudhuri equation.

Very close to the singularity string-scale effects gain prominence and we may need to correct the bulk description using $\alpha'$ corrections \cite{Zigdon:2024ljl, Martinec:1994xj} (see also \cite{Dodelson:2025jff, Afkhami-Jeddi:2025wra}). It would also be interesting to diagnose aspects of near-singularity behaviour using an asymptotic measurements \cite{Frenkel:2020ysx, DeClerck:2023fax, DeClerck:2025mem}.

\section{Conclusion}
\subsection{Discussion}
We firstly summarize our work. We provide a description of the black hole interior which can be obtained using by acting with integral transforms boundary correlators. This take the exterior measurement using four-point correlator and transforms it to describe local flat-space like physics about a bulk point in the analytically continued interior. The broader theme is similar to \cite{Afkhami-Jeddi:2025wra} where a Fourier transform on the two-point function tunes it to receive contributions from reflecting geodesics in the interior. 

Our procedure necessitates inclusive amplitudes about the interior bulk point. To look inside black holes, we employ scattering experiments with general operator-ordering prescriptions. Standard in-out experiments do not work well here as the end products can fall inside the horizon. However inclusive amplitudes offer more choices where we do not require additional data from regions behind the horizon. We construct a radar-like experiment as well as more general out of time ordered correlators. Specific to the thermofield doubled state, starting with a four point otoc we also arrive at the interior bulk point. On line of \cite{Caron-Huot:2025hmk}, our work demonstrates other instances where shooting directed wavepackets using the smeared CFT correlator factorizes to a universal formula giving us flat-space like scattering amplitudes. We analytically continue these exterior experiments with insertions on (multiple) timefolds, such that all measurement data can be recovered using only one side. 

In the presence of Killing symmetries such as in empty AdS, we can search for the bulk points using four point correlator. However in general geometries, sharply singling out a bulk point is difficult as the bulk wavefront emerging from a generic boundary deformation is not localized along a particular direction. Consequently the conformal correlator picks up contributions from all possible kinematically allowed bulk points. Locating a bulk point thus involves a $d+2$ dimensional correlator to compensate for the dimensionality. Additionally we also need to satisfy the local flat space bulk momenta constraint at the point $X$, i.e. $\sum_i P_i =0$ \cite{Maldacena:2015iua}. 

This leads to another central ingredient, i.e., directed wavepackets which are introduced by integrating the boundary position-space operators against a Fourier transform that has a Gaussian weight. We work with Wightmann functions where in the high energy limit, the evolution of the wavepacket is governed by a geodesic approximation of the bulk-to-boundary propagator which goes from the boundary to the bulk point. We can think about the geodesics as introducing a dressing for the flat space processes. From this perspective, we firstly consider a flat-space process where locally $\sum_i P_i =0$ is ensured. Next, we simply dress the process to the AdS boundary using the geodesics. The first two orders in the WKB expansion helps us understand features involving physical optics. 

 Using these ingredients, we provide a dictionary for local emission and absorption operators to smeared boundary operators. Consequently the boundary correlator reduces to a simple factorization formula.  As stressed in \cite{Caron-Huot:2025hmk}, the factorization formula is derived by ignoring interactions outside the neighborhood of the bulk point such as long-range forces. Nevertheless the formula captures the essential physics about the bulk point and is a good approximation to the full scattering problem.  The scattering contribution that we obtain is not really the S-matrix but its hard part.

As we state earlier, the backward Euclidean time evolution due to the transform taking oscillators from the left boundary to the right boundary creates geometries that violate KSW criteria. However since both the setup as well the operations are concrete, it is no surprise that we still obtain physically meaningful conclusions. More generally geometries that violate such stability criteria may not be sick, see for instance recent discussions about global symmetries \cite{Bah:2022uyz}, bra-ket wormholes \cite{Fumagalli:2024msi} and saddles of the static patch spectrum \cite{Chakravarty:2025sbg} in de Sitter.

While the four-point function seems a sharp probe of bulk locality in the interior away from the singularity, close to the singularity the smaller time scales are probably better suited to probe the same, rather than our present experiments with four point correlators. More explicitly, this is so since the Wightman two-point function on second sheet has an actual contribution from the black hole singularity at complex time $t = \lc \pm \frac{\beta}{2} + \frac{\ic \beta}{2}\rc$.

\subsection{Future directions}

We now discuss about some important points, list future directions and provide connections to existing literature. 

\textbf{Two-sided correlations:} Presently we utilized a smearing transform which allows us to create a local experiment in the black hole interior using exterior scattering data. However these asymptotic measurements does not shed light on whether there actually exist dragons behind the horizon unless we send an infalling observer in.

With another asymptotic boundary on the left, we may be able to answer this question better. One may ask whether a suitable analytic continuation of the two-sided four-point correlator can provide a sharp measurement of local features in the interior. This can then be tested against a deformation of the thermofield doubled state which has a localized shock or a shell in the future interior. 

\paragraph{Collapse and instabilities near singularity:} Recently progress has been made towards understanding local evolution close to Schwarzchild and (more stable) BKL singularities, including deformations to general Kasner behaviour \cite{Frenkel:2020ysx, DeClerck:2023fax, Caceres:2023zft, Bueno:2024fzg, DeClerck:2025mem}. We believe a systematic boundary understanding of these aspects in terms of holographic correlators is crucial to incorporate dynamical collapse scenarios and its effects on the Schwarzchild interior. The black hole interior is quite unstable, we expect infalling matter to drastically alter the dynamics upon going away from our present probe description. There also have been studies of collapsing scenarios in controlled models \cite{Anous:2016kss, Anous:2017tza, Dhar:2018pii}. It would be great to connect these bulk interior descriptions concretely to asymptotic measurements using boundary correlators. In this light, we can probably start by addressing a much simpler question: how is a general wormhole geometry dual to a generic entangled state of two CFTs imprinted on the four-point correlator with support on both sides? 

\textbf{Recovering infalling diaries:}  It might be useful revisit the Gao-Jafferis-Wall protocol in the light of our present discussion which allows us to perform experiments on some part of the final state \cite{Gao:2016bin, Maldacena:2017axo}. Here we can represent the double-trace deformation can be written as a smeared four-point otoc, it would be interesting how naturally our present otoc discussion is connected to this. More generally, the Hayden-Preskill protocol \cite{Hayden:2007cs} can be represented as a four-point correlator \cite{Yoshida:2017non} in simple models, it would be interesting to connect this to our present discussion of how the black hole interior is imprinted upon the thermal four-point correlator. 
 
 A complementary way which might also be useful is to directly track the experience of an infalling observer by measuring physical quantities using worldline correlators along the lines of \cite{Sivaramakrishnan:2024ydy, Sivaramakrishnan:2025srr, Bohra:2025mhb}. Related to this, see also \cite{Kaushal:2025wbn} which utilizes a boundary measurement to provide a sharp signature of an infalling geodesic crossing the horizon. 

\paragraph{Thermal correlators using CFT:} While we utilized the cylinder to plane map in 2d CFT to understand features of the planar BTZ interior, performing a useful analysis in higher dimensions is hard. Nevertheless it might be useful to aim at obtaining the flat space limit in both the interior and the exterior geometries using a thermal four-point correlator description. This would be useful to verify the universal aspects of flat-space limit as well as quantify subleading corrections to the same.
Even in our 2d CFT example we restricted to the planar BTZ case, it might be useful to utilize the monodromy methods of \cite{Fitzpatrick:2016mjq} to work out the spherical BTZ case.

On a slightly different note, the framework of holographic cameras work nicely using the conformal Regge theory technology, it might be useful to understand whether similar analysis can be performed at the level of multi-Regge theory \cite{Costa:2023wfz}.

 \textbf{Bulk measurements (and reconstruction?):} Our dictionary to obtain bulk oscillator modes requires four wavepackets smeared over  points on the past and the future boundary hyperboloids. Here the kinematics of the past bulk lightcones from a bulk point project out loci on the boundary, where taking $t_R \to -t_L - {\ic \beta \over 2}$ relates the future hyperboloid on right boundary to the past hyperboloid on the left boundary. As we stress earlier, the interior measurements are a relatively straightforward extension of exterior physics. Simple operations like taking derivatives on the boundary hyperboloids tell us the bulk metric upto a conformal factor \cite{Engelhardt:2016wgb, Caron-Huot:2025she}. To fix the conformal factor, we need to perform a measurement such as the flat space amplitude described here. The overall intensity of the resulting factorization formula has a $\sqrt{-g}$ in front, which completely fixes the conformal factor. 

 These measurements are simpler compared to causal bulk reconstruction, since we require only four wavepackets compared to an entire subregion \cite{Hamilton:2005ju, Hamilton:2006az} (see \cite{Kajuri:2020vxf} for a review, and \cite{Hamilton:2006fh, Papadodimas:2012aq} for extensions to interior). These measurements are also simpler compared to entanglement wedge reconstruction, which also requires access to larger boundary subregions \cite{Ryu:2006bv, Czech:2012bh, Jafferis:2015del, Dong:2016eik}.


\paragraph{Universality of the flat-space limit:} Finally, regarding measurements, we stress the universality as in \cite{Caron-Huot:2025hmk, Caron-Huot:2025she}: there exists a flat-space like factorization formula irrespective of the bulk point's location which is indicative of the equivalence principle. Consider the exterior experiment in an ambient spacetime where we send two electrons into a CFT from an ambient flat spacetime \cite{Caron-Huot:2025hmk}. Even basic bulk scattering is non-trivial from the CFT perspective: corresponding to early time creation operators on boundary appears, expanding ripples of positive energy following \cite{Hofman:2008ar} go out on the boundary from the past hyperboloid (and expanding ripples of negative energy for absorption operators). Similar ripples emergy from locations on late time hyperboloid. Using these we know a scattering event has taken place in the bulk, it would be interesting to understand the boundary picture here.

The interior case is even more striking: upon performing our transformation in \eqref{finform} on the exterior otoc experiment, we find the exterior experiment becomes equivalent to shooting in electrons which are two expanding ripples into the right CFT which upon bulk scattering leads to ripples emerging in the left CFT but opposite in time (since the left CFT time reverses). Thus from the perspective of the left CFT, when a bulk scattering event happens in the interior, we see contracting ripples and then electrons emerge into the ambient spacetime, which is completely crazy! We feel that understanding the basic boundary physics which underlies these aspects could shed further light on why gravity mysteriously arises from a non-gravitating system.

\section*{Acknowledgements}
I thank Simon Caron-Huot for collaborating in the early stages, for very insightful discussions, and extremely helpful advice which have greatly improved the work. I also thank Nima Afkhami-Jeddi, Tarek Anous, Chandramouli Chowdhury,  Keshav Dasgupta, Marine De Clerck, Sean Hartnoll, Diksha Jain, R Loganayagam, Alex Maloney, Viraj Meruliya, Keivan Namjou, Chintan Patel,  Priyadarshi Paul, Suvrat Raju, Rodolfo Russo, Shivam Sharma, Kaustubh Singhi, Akhil Sivakumar, Joao Vilas Boas, Aron Wall and Yoav Zigdon for various helpful suggestions, related discussions, as well as for probing questions about related topics that allowed us to understand the subject much better. Our work is supported by the National Science and Engineering Council of Canada (NSERC) and the Canada Research Chair program, reference number CRC-2022-00421. 

\appendix
\section{Wavepackets and bulk points}
We put in discussions regarding wavepackets and bulk points that supplement the discussion in the main text.
\subsection{Wavepacket propagation in the bulk}
\label{sec:wavepackets}

We summarize important formulae from \cite{Caron-Huot:2025hmk} here.
\paragraph{Wave equation:} Our objective here is to find a simple expression for the bulk-to-boundary propagator in terms of a phase. In the presence of symmetries along the transverse directions, the transverse bulk momenta are related to the boundary momenta as $P_\mu = p_\mu$.  We thus work with the plane wave ansatz $\Phi_p(X) = \ep^{\ic p_\mu X^\mu} \phi_p(z)$, the equation of motion is given by:
\begin{equation}\label{eq:eom-scalar-pbh}
    z^{d+1} \partial_z \qty(z^{-d+1} f(z) \partial_z \phi_p(z)) - \qty(z^2p^2 + m^2) \phi_p(z) = 0, \qquad p^2=\eta^{\mu\nu}p_\mu p_\nu.
\end{equation}
In the high energy limit, we evaluate the propagators using the WKB approximation. We are interested in the physical optics limit, where we need only the leading and the sub-leading orders in WKB approximation. 
\beq \label{sol1}
\phi^\pm_{p}(z)  \propto  \sqrt{\frac{z^{d-1}}{A(z)P_z}}
    \exp(\pm \ic \int_0^z \diff z \, P_z + \mathcal{O}\qty(\frac{1}{P_z}))
\eeq
Here the radial momentum $P_z$ is the large parameter in the WKB approximation.
\beq \label{Pzee}
 P_z = \sqrt{- \frac{g^{\mu\nu}}{g^{zz}} p_\mu p_\nu} = {1 \over f(z)} \sqrt{{E^2 - \mathbf{p}^2 f(z)}},
\eeq
We determine the normalization constant $\mathcal{C}^\pm_{\Delta,p}$ by matching the WKB solution with the exact empty AdS solution at the asymptotic boundary.
\beq \label{normalization1}
  \mathcal{C}^\pm_{\Delta,p} = \ep^{\pm \ic \theta_\Delta}\sqrt{\frac{2 \pi^{\frac{d}{2}+1}}{\Gamma(\Delta) \Gamma\qty(\Delta - \frac{d}{2} + 1)}}\qty(\frac{\sqrt{-p^2}}{2})^{\Delta - \frac{d}{2}}, \qquad \qquad \theta_\Delta = \frac{\pi}{4} \qty(d - 1-2\Delta)
\eeq
The $\pm$ signs in \eqref{sol1} denote the radially ingoing and outgoing modes respectively in the bulk. These choices arise since the symmetric Wightman propagator can be expressed as a sum of the time-ordered and anti-time-ordered correlators:
\beq \label{wight}
\begin{split}
\Phi_{x,p,\sigma}(X) 
=  {\rm T} \lc \Phi_{x,p,\sigma}(X) \rc + {\bar{\rm T}} \lc \Phi_{x,p,\sigma}(X) \rc
\end{split}
\eeq
The above identity allows the boundary operator to have support for bulk points both in the causal past and the causal future.
\paragraph{Finite width wavepackets:} Here we compile some formulae from \cite{Caron-Huot:2025hmk} for a self-contained discussion. Deviating from the plane wave basis to our case with finite smearing leads to the following relation between the bulk modes and the boundary smeared operators:
\begin{equation}\label{eq:dictionary wavepacket}
O^\dagger_{x,p,\sigma}
    \simeq
    \int \frac{\diff^{d+1}P}{(2\pi)^d\sqrt{-g}}\delta\qty(P^2) \,
    \bar{\psi}^{\rm bulk,+}_{x, p, \sigma; X}(P)
    \, a^\dagger_{X,P},
\end{equation}
when $O^\dagger$ is to the past of the bulk point $X$ (a similar formula involving the bulk oscillator $b^\dagger_{X,P}$ holds if $O^\dagger$ is to the future of the bulk point $X$ ) Note here that we have a bulk wavepacket in the momentum space, where the broadening function $ \tilde{\psi}^{\rm bulk}_{x,p,\sigma; X}(P+\delta P)$ takes the following form:
\begin{equation} \label{psi bulk good}
    \tilde{\psi}^{\rm bulk,\pm}_{x,p,\sigma; X}(P+\delta P) \approx \frac{\mathcal{C}^\pm_{\Delta,p}\sqrt{\det2\pi\sigma}}{\sqrt{\mathcal{D}(p;X)}} \exp(\ic S(p;X) - \ic \Delta X^M\delta P_M-\frac12 \Sigma^{MN} \delta P_M\delta P_M),
\end{equation}
where we have defined,
\begin{equation} \label{psi bulk X Sigma}
    \Delta X^M = \Delta x^\mu\frac{\partial k_\mu}{\partial P_M},\qquad
    \Sigma^{MN} = \Sigma^{\mu\nu}\frac{\partial k_\mu}{\partial P_M}\frac{\partial k_\nu}{\partial P_N}, \qquad    \frac{\partial k_\mu}{\partial P_M} = \qty[\begin{pmatrix}\frac{\partial P_M}{\partial p_\mu}& n_M\end{pmatrix}^{-1}]_{{\rm top\,} d {\rm\,rows}}.
\end{equation}
Let us define the measure as
\begin{equation} \label{prodpsi}
\begin{split}
\qty[\prod \diff P_j\,\psi_j]=
    \prod_{j=1}^{m+n} \frac{\diff^{d+1}P_j}{(2\pi)^d\sqrt{-g}} \delta\qty(P_j^2)\theta(P^0)\,
 \tilde{\psi}^{\rm bulk}_{x_j, p_j, \sigma_j; X}(P_j).
 \end{split}
\end{equation}
Will all things in place, the $n$-point boundary correlator when integrated against smeared Fourier transforms factorizes to a flat-space inclusive amplitude given below:
\begin{equation} \label{finexp}
\begin{array}{c}
\Pi_{\{\sigma_i,p_i,q_i\}}(x_i) \equiv\langle {\Psi} \,| \, \text{(product of $O_{x,p,\sigma}, O^\dagger_{x,p,\sigma}$'s and $O_{y,q,\sigma}, O^\dagger_{y,q,\sigma}$'s)} \, |\, {\Psi} \rangle \\[.5em]
\updownarrow \\[.5em]
 \int \prod_{i=1}^n {\rm d} (\delta P_i) \, \psi_i(\delta P_i) \, 
\langle \Omega \,  | \,  \text{(product of $a$, $a^\dagger$, $b$, $b^\dagger$'s)} \, | \, {\Omega} \rangle
\end{array}
\end{equation}
where $\psi_i(\delta P_i)$ is a bulk Gaussian factor and $a_{X,P}| \, {\Omega} \rangle = 0$.

The above expression is the factorization formula which relates the smeared correlator to the bulk scattering amplitude. Importantly, the time-ordering of the flat space correlator remains the same as the CFT correlator. Upon substituting the expression for the flat space amplitude in the above expression, we indeed obtain the formula given in \eqref{rep2}.

\subsection{More on bulk points}
\label{sec:morebp}
We explain why the singularity arises from the vacuum using an explicit D-function analysis as well as briefly sketch the Mellin space description.
\subsubsection{An explicit D-function example} 
As a simple example the above general behaviour of the bulk point singularity in \eqref{div} can be explicitly seen by studying the Witten diagrams, which are represented in terms of D-functions. As an example, let us take the Witten diagram where we set all $\Delta_i =1$:
\beq
{\rm D}_{\Delta_i =1}(z, \bar{z}) = \frac{(z \bar{z})}{\bar{z}-z} \left(2 \text{Li}_2(z)-2 \text{Li}_2(\bar{z})+\log \left(\frac{1-z}{1-\bar{z}}\right) \log (z \bar{z})\right)
\eeq
Taking one of the cross-ratios around the two branch-cuts corresponding to Fig \ref{redpath} gives us the bulk-point discontinuity.
\beq
{\rm D}_{\Delta_i =1}^\curvearrowright(z, \bar{z}) =\pm \frac{4 \pi ^2 z \bar{z}}{z-\bar{z}}
\eeq
Here the $\pm$ signs depend on the relative directions while crossing the branch cuts at $0$ and $1$ (and also on which cross-ratio we have chosen). 

Bulk scattering also takes place when the cross-ratios approach $z \to \bar{z}$ across the branch cut as Re$ z, \bar{z} <0$. We again take $z$ around $z=1$ while again keeping $\bar{z}$ fixed. From this configuration letting both cross ratios hitting the bulk point discontinuity at Re$z,\bar{z} < 0$ gives us:
\beq
{\rm D}_{\Delta_i =1}^\curvearrowright(z, \bar{z}) = \pm \frac{4 \ic \pi  z \bar{z}}{z-\bar{z}} (\log (z)+\log (\bar{z})) \approx \mp \frac{4 \pi ^2 z \bar{z}}{z-\bar{z}}
\eeq
where $\pm$ indicates whether the cross ratios went (counter-)clockwise.
The D-functions with $\Delta_i \neq 1$ are related to $\Delta_i =1$ using derivatives \cite{Dolan:2000ut, Gary:2009ae} which decide the value of $n$ in \eqref{div}.

\subsubsection{Mellin space} It is useful to understand the appearance of the singularity and determine the factor $n$ in \eqref{div} using the Mellin space description. The relationship between the Mellin amplitude $M$ and the flat-space scattering amplitude $\mathcal{M}$ is given by the integral transform \cite{Penedones:2010ue, Fitzpatrick:2011hu}:
\beq \label{eq:FlatToMellin}
\mathcal{M}(s_{ij})=R_{\rm AdS}^{\frac{n(d-1)}{2}-d-1} \Gamma\lc \frac{\Delta_\Sigma-d}{2}\rc \int_{-\ic \infty}^{\ic \infty} \frac{\diff \alpha}{2\pi \ic} \, e^{\alpha} \alpha^{\frac{d-\Delta_\Sigma}{2}}M\lc \delta_{ij}=-\frac{R_{\rm AdS}^2}{4\alpha}s_{ij}\rc.
\eeq
To analyze the bulk point singularity, we need to express the Mellin amplitude $M$ in terms of the flat-space amplitude $\mathcal{M}$, i.e. the inverse of \eqref{eq:FlatToMellin}. Recognizing \eqref{eq:FlatToMellin} as an inverse Laplace-like transform, we can invert it to obtain:
\beq \label{eq:MellinToFlat}
M(\delta_{ij}) \sim \mathcal{N} \int_0^\infty \diff \beta \, e^{-\beta} \beta^{\frac{\Delta_\Sigma - d}{2} - 1} \, \mathcal{M}\lc s_{ij} = -\frac{4\beta}{R_{\rm AdS}^2}\delta_{ij}\rc.
\eeq
where $\mathcal N$ can be fixed using the inverse Laplace transform. For the specific case of the four points, the correlator is defined as the following transform:
\beq
\mathcal{G}(u, v) = \int_{-\ic\infty}^{\ic\infty} \frac{\diff s \diff t}{(4\pi \ic)^2} \, u^s v^t \, \Gamma^2(-s) \Gamma^2(-t) \Gamma^2\lc \frac{\Delta_\Sigma - d}{2} + s + t \rc \, M(s, t),
\eeq
where $\Delta_\Sigma = \sum_i \Delta_i$. In the high-energy limit  with large Mellin variables $s,t$ where we suitably avoid the Gamma function poles, the conformal dimensions $\Delta_i$ become negligible and appear in subleading corrections to the saddle-point phase. The integral is dominated by the Jacobian of the mapping from Mellin space to the impact parameter space. The correlator $\mathcal{G}(z, \bar{z})$ basically scales as \eqref{div} with $n=d-3$.

\section{WKB beyond the geodesic limit }
\label{energy-expansion}

We perform a WKB analysis of the differential equation for the bulk-to-boundary propagator in backgrounds with temporal and transverse symmetries described by \eqref{eq:pbhmetric}. The direct way to compute corrections to the bulk to boundary propagator is by performing the semiclassical expansion to all orders of the solution using the energy expansion.  Given the equation of motion,
\beq \label{eomphi}
 z \left(\phi '(z) \left(z f'(z)-(d-1) f(z)\right)+z f(z) \phi ''(z)\right)+\phi (z) \left(\frac{\omega ^2 z^2}{f(z)}-{\bf p}^2 z^2-m^2\right) = 0
 \eeq
we write the propagator using the form $\psi(z) = \exp  A(z) $, where we substitute the following expansion for the solution:
 \beq
A(z) = \omega \sum_{j=0}^\infty {A_{j+1}(z) \over \omega^j}
\eeq
We are interested in the exponent using the high-energy expansion. Upto the first four orders this takes the form.
\beq
\phi_p(z) \sim \exp \left(\omega  A_1(z)+A_2(z)+\frac{A_3(z)}{\omega }+\frac{A_4(z)}{\omega ^2}\right)
\eeq
Substituting the above expansion into the equation of motion, we obtain the following differential equations at each subsequent order in the energy expansion.
\beq
\begin{split}
    &\frac{z^2}{f(z)}-z^2 f(z) A_1'(z){}^2  = 0 \\
    &\ic z \left(A_1'(z) \left(f(z) \left(2 z A_2'(z)-d+1\right)+z f'(z)\right)+z f(z) A_1''(z)\right)  = 0\\
    &z f(z) \left(-(d-1) A_2'(z)+z \left(A_2''(z)+2 \ic A_1'(z) A_3'(z)\right)+z A_2'(z){}^2\right)+z^2 A_2'(z) f'(z)-{\bf p}^2 z^2-m^2  =0 \\
    &z \left(z A_3'(z) f'(z)+f(z) \left(A_3'(z) \left(2 z A_2'(z)-d+1\right)+z \left(A_3''(z)+2 \ic A_1'(z) A_4'(z)\right)\right)\right)=0
\end{split}
\eeq
Starting with the first equation for $A_1(z)$, we can solve all the differential equations one by one by substituting the solution at the  previous orders. We obtain:
\beq
\begin{split}
A^\pm_1(z) &= \pm \ic \int \frac{1}{f(z)} \, dz \\
A_2(z) &= \frac{1}{2} (d-1) \log (z) \\
A_3(z) &= \pm \ic \int \frac{ \left(d^2-1\right) f(z)-2 (d-1) z f'(z)+4 \left({\bf p}^2 z^2+m^2\right)}{8 z^2} \, dz \\
A_4(z) &= \frac{f(z) \left(\left(d^2-1\right) f(z)-2 (d-1) z f'(z)+4 \left({\bf p}^2 z^2+m^2\right)\right)}{16 z^2}\\ 
\end{split}
\eeq
The above expressions have constants of integration appearing at each order in frequency. The integration constant at the leading order is fixed using the exact value of the two point function near the boundary, where the geometry is empty AdS (see \eqref{normalization1} for exact expression). Here the leading near-boundary solution is given by Hankel functions, which gives rise to a normalization constant and a phase shift \cite{Caron-Huot:2025she}. To obtain the subleading corrections, we need to match the subleading WKB with the subleading near-boundary correction. For instance, the sub-leading solution can be extracted using Green function techniques upon using the leading Hankel solution. We do not record the calculation here, but it is solvable using Mathematica.

A general feature of the higher order terms is that they do not have branch cuts, but have regular poles at the boundary and the black hole singularity.
\section{Unruh modes in two dimensional Rindler space }
 \label{2drindler}
We work with flat space in two dimensions to demonstrate properties of Rindler modes and their relation to Minkowski modes. The two-dimensional Minkowski metric is given by:
\beq
{\rm d}s^2 = {\rm d}t_M^2 - {\rm d}x_M^2.
\eeq
We can rewrite the coordinates using $U = t_M-x_M,$ and $ V = t_M+x_M$. In the region $x_M>\abs{t_M}$, we define the Rindler coordinates for Rindler observers with unit acceleration $a=1$:
\beq
\begin{split}
    U = -  \, \exp \qty(- {2 \pi \over \beta} \,u), \qquad  V = \, \exp \qty({2 \pi \over \beta} \,v).
\end{split}
\eeq
where $u \equiv t-x, \, v \equiv t+x$, where $x$ and $t$ denote the Rindler space and time coordinates respectively. Note here that the Rindler temperature for unit acceleration is $\beta  =2\pi$, but we explicitly keep $\beta$ to demonstrate the analogy with the black hole case.

We call this region as R as depicted in Figure \ref{fig:rindler}. The coordinates on  right (future) horizon is given by $U = 0$, while the left (future) horizon coordinate is $V = 0$. The metric in these new coordinates takes the conformal form
\beq
{\rm d}s^2 = e^{\lc 4\pi \over \beta \rc x}  \lc {\rm d}t^2 - {\rm d}x^2\rc,
\eeq
 One can define other quadrants similarly by changing the signs in front of the exponential analogously to the black hole case. Going from Region R to F by a future directed infalling geodesic is implemented by
 \beq
t \to t - {\ic \beta \over 4}, \qquad x \to x + {\ic \beta \over 4}
 \eeq
 while going from Region F to L is further implemented by
 \beq
t \to t - {\ic \beta \over 4}, \qquad x \to x - {\ic \beta \over 4}
 \eeq
 which gives us a total change from Region R to Region L
 \beq
t \to t - {\ic \beta \over 2}, \qquad x \to x.
 \eeq

\subsection*{Rindler mode expansions}
Since the modes are conformally flat, we can define a mode expansion similar to the Minkowski expansion. For convenience, we use the dispersion relation to write the spatial momentum $p$ in terms of the frequency $\omega$. For region R, this takes the form:
\beq
\phi_{\rm R}(u,v) = \int \frac{{\rm d}\omega}{\sqrt{\omega}} \lc f_{\omega} e^{-\ic\omega u} + g_{\omega} e^{-\ic\omega v} + \text{h.c.}\rc.
\eeq
The mode expansion takes a similar form for Region L, which is given by
\beq
\phi_{\rm L}(u,v) = \int \frac{{\rm d}\omega}{\sqrt{\omega}} \lc \tilde{f}_{\omega} e^{\ic\omega u} + \tilde{g}_{\omega} e^{\ic\omega v} + \text{h.c.}\rc.
\eeq
Note that the exponential signs here are different since the Rindler time $t$ runs in the opposite direction in Region L compared to Region R. Consequently, on the slice at $t=0$, we have the following mode expansion:
\beq \label{thrindd}
\phi(u,v) = \int \frac{{\rm d}\omega}{\sqrt{\omega}} \lc f_{\omega} e^{-\ic\omega u} + g_{\omega} e^{-\ic\omega v}+ \tilde{f}_{\omega} e^{\ic\omega u} + \tilde{g}_{\omega} e^{\ic\omega v} + \text{h.c.}\rc.
\eeq

\subsection*{Unruh modes and relation to Rindler modes}
Our goal here is to determine relation between Minkowski and Rindler modes. To do this, we instead define Unruh modes and use them to find the Bogoliubov transformation conveniently between the Rindler and the Unruh modes. It turns out that the Unruh and Minkowski vacuum are essentially the same. This is quite convenient, for instance, this considerably simplifies the calculation of Rindler-to-Minkowski Bogoliubov coefficients.

The Unruh mode is defined in the two different regions as follows:
\beq \label{thunr}
\begin{split}
    U_{\rm U}(u) = e^{-\ic\omega u}, \quad \text{Region L}; \qquad
    U_{\rm U}(u) = e^{\frac{\beta \omega}{2}} e^{-\ic\omega u}, \quad \text{Region R} 
\end{split}
\eeq
which can be essentially written as $U_{\rm U}(u) =  U^{\frac{\ic\omega \beta}{2 \pi}}$. Note that $U^{\frac{\ic\omega \beta}{2 \pi}}$ has a branch cut, which we can choose in the upper half plane. Apart from this $U$ does not have any singularities in the lower half plane. Consequently one can analytically continue $U$ in the lower half plane from Region L to Region R. This gives rise to the extra factor of $e^{\frac{\beta \omega}{2}} $ in eqn \eqref{thunr}.

Since Unruh modes are analytic in the lower half-plane, it is easy to see that the modes satisfy the following property
\beq
\int_{-\infty}^{\infty}{\rm d}U \, U^{\frac{\ic\omega \beta}{2 \pi}} \, e^{-\ic\omega' U} = 0, \qquad \text{for} \qquad \omega' >0
\eeq
as the integral can be continued analytically in the lower half-plane and is zero. Thus the Unruh mode has only positive Minkowski frequencies. 

Using the $u$ coordinate one can similarly define the other Unruh $\tilde{U}_{\rm U}(u)$ mode as
\beq
\begin{split}
    \tilde{U}_{\rm U}(u) = e^{\ic\omega u}, \quad \text{Region L}, \qquad
    \tilde{U}_{\rm U}(u) = e^{-\frac{\beta \omega}{2}} e^{\ic\omega u}, \quad \text{Region R} 
\end{split}
\eeq
where we have chosen the branch cut in the lower half-plane, and hence analytically continued through the upper-half plane. We can similarly define the Unruh mode using the $v$ Rindler mode, which is given by:
\beq
\begin{split}
    V_{\rm U}(v) = e^{-\ic\omega v}, \quad \text{Region L} , \qquad
    V_{\rm U}(v) = e^{\frac{\beta \omega}{2}} e^{-\ic\omega v}, \quad \text{Region R} 
\end{split}
\eeq
while the conjugate Unruh mode $\tilde{V}_{\rm U}(v)$ is given by
\beq 
\begin{split}
    \tilde{V}_{\rm U}(v) = e^{\ic\omega v}, \quad \text{Region L}, \qquad  \tilde{V}_{\rm U}(v) = e^{-\frac{\beta \omega}{2}} e^{\ic\omega v}, \quad \text{Region R} .
\end{split}
\eeq
The mode expansion of the scalar field in the Unruh modes is given by
\beq
\phi(u,v) = \int \frac{{\rm d}\omega}{\sqrt{\omega}} \lc d_{\omega} U_{\rm U}(u) + \tilde{d}_{\omega} \tilde{U}_{\rm U}(u)+ e_{\omega} V_{\rm U}(v) + \tilde{e}_{\omega} \tilde{V}_{\rm U}(v)+ \text{h.c.}\rc.
\eeq
Since there is no mixing of positive and negative frequencies, both the Unruh vacuum and the Minkowski vacuum are the same, i.e. $d_{\omega} | \Omega_M\rangle = \tilde{d}_{\omega} | \Omega_M\rangle =0$. From the Rindler oscillators defined in eqn \eqref{thrindd}, we can simply read off the Rindler to Unruh coefficients:
\beq
\begin{split} \label{thrtu}
    d_{\omega} = \frac{ f_{\omega} - e^{-\frac{\beta \omega}{2}} \tilde{f}_{\omega}^{\dagger}}{ e^{\frac{\beta \omega}{2}} - e^{-\frac{\beta \omega}{2}}} \quad \text{and} \quad \tilde{d}_{\omega} = \frac{f^\dagger_{\omega} - e^{\frac{\beta \omega}{2}} \tilde{f}_{\omega}}{ e^{-\frac{\beta \omega}{2}} - e^{\frac{\beta \omega}{2}}}
\end{split}
\eeq
It can be checked that these imply $d_{\omega} | \Omega_M\rangle = \tilde{d}_{\omega} | \Omega_M\rangle =0$. These are the same equations we utilized in the black hole case. Using these Bogoliubov coefficients, one can write the Minkowski vacuum in terms of the Rindler vacuum as follows:
\beq
\ket{\Omega_M} \propto \exp \lc \sum_{\omega} e^{-\frac{\pi \omega}{a}} f^{\dagger}_{\omega} \tilde{f}^{\dagger}_{\omega} + g\text{-terms}\rc \ket{\Omega_{R,L}}
\eeq
One can expand this state which results in the well-known property of Minkowski vacuum that it is a thermofield doubled state when written in terms of Rindler modes: 
\beq
\ket{\Omega_M} = {1 \over \sqrt{Z(\beta)}}\sum_E e^{-\frac{\beta E}{2}} \ket{E,E}
\eeq
where $E = \sum \omega \, n_{\omega}$ is the energy of the state. Note that if we sum over one of either the regions L or R, we get the canonical density matrix for the other region, i.e.
\beq
\rho(E) = \frac{e^{-\beta E}}{Z(\beta)}
\eeq
For a similarly motivated summary of the general dimensional case, refer to \cite{Birrell:1982ix, Papadodimas:2012aq, Raju:2018zpn, Chakraborty:2021rvy}.

\bibliographystyle{JHEP}
\bibliography{citation}

\end{document}